\documentclass{aa}

\usepackage[]{graphics}
\usepackage[]{psfig}
\usepackage[]{epsfig}
\usepackage[]{longtable}
\include{epsf}
\begin{document}

\title{Distance to the Centaurus cluster and its subcomponents from surface brightness 
fluctuations\thanks{Based on observations obtained at the European Southern Observatory,
    Chile (Observing Programme 67.A--0358).}}

\author {Steffen Mieske \inst{1,2} \and Michael Hilker \inst{1}}

\offprints {S.~Mieske}
\mail{smieske@astro.puc.cl}

\institute{
Sternwarte der Universit\"at Bonn, Auf dem H\"ugel 71, 53121 Bonn, Germany
\and
Departamento de Astronom\'\i a y Astrof\'\i sica, P.~Universidad Cat\'olica,
Casilla 104, Santiago 22, Chile
}

\date {Received 19 May 2003 / Accepted 28 July 2003}

\titlerunning{Distance to the Centaurus cluster}

\authorrunning{S.~Mieske \& M.Hilker}

\abstract{We present $I$-band Surface Brightness Fluctuations (SBF) measurements for 15 early 
type galaxies (3 giants, 12 dwarfs) in the central
region of the Centaurus cluster, based on deep photometric data in 7 fields obtained
with VLT FORS1 and with very good seeing.
From the SBF-distances to our sample galaxies we determine the distance of the 
Centaurus cluster to be 41.3 $\pm$ 2.1 Mpc (33.08 $\pm$ 0.11 mag). This places 
the Centaurus cluster at about the same distance as the ``Great Attractor''. 
We find a distance
difference of 0.27 $\pm$ 0.34 mag between the two subcomponents Cen30 and Cen45, 
ruling out that both components 
are separated by their
Hubble flow distance.
A distance difference of 0.48 $\pm$ 0.21 mag is found between 
the central galaxies NGC 4696 (Cen30) and NGC 4709 (Cen45) of both components, supported by the different
turn-over magnitudes of their respective globular cluster systems. This suggests that 
Cen45 is falling into but has not yet reached Cen30, supporting the idea of a large scale filament along 
the line
of sight towards Centaurus (Churazov et al. \cite{Churaz99}). $H_{\rm 0}$=83.0 $\pm$ 8.3 km/s/Mpc 
is obtained for 
our Cen30 sample 
taking into account the peculiar motion of the Local Group into the direction of the Centaurus cluster.
This value of $H_{\rm 0}$ corresponds to a much smaller Hubble flow distortion in the
direction of Centaurus than determined
by Tonry et al. (\cite{Tonry00}), implying that the GA mass estimate
by Tonry et al. may be too high and/or that the Centaurus cluster falls into the GA almost
perpendicularly to the line of sight.
As our mean single measurement error is very
close to the measured distance scatter of the investigated galaxies, we can only derive an 
upper limit of $\pm$ 10 Mpc radial extension for the Centaurus cluster, 
corresponding to a five times larger radial than tangential extension.
No evidence for an infall pattern into the Great Attractor is found within the uncertainties
for the 11 galaxies
with measured redshifts.}

\maketitle


\section{Introduction}
\label{Introduction}
\subsection{Attractors}
In the nearby universe, peculiar galaxy velocities with respect to the Hubble 
flow caused by clumpy matter distribution can constitute a significant fraction 
of the total radial velocity, which introduces a bias into the measurement
of the Hubble constant $H_{\rm 0}$ 
if not corrected for. 
The 
two most prominent and best studied nearby matter concentrations causing 
deviations from the Hubble flow are the Virgo Attractor (VA) at about 17 Mpc
distance and the 
``Great Attractor'' (GA) at about 43 Mpc distance close to the Centaurus
Cluster of galaxies (Dressler et al. \cite{Dressl87}, Tonry et al. \cite{Tonry00}, in the 
following referred to as {\it SBF II}, as abbreviated by the Tonry group). The gravitational 
pull of the VA is about 
150 $\pm$ 50 km/s, of the GA it is about 300 $\pm 100$ km/s ({\it SBF II}).\\
The three-dimensional position of the VA
was identified with the Virgo cluster of galaxies 
more than twenty years ago (e.g. Schechter \cite{Schech80}, Yahil et al. \cite{Yahil80}, 
Tonry \& Davis \cite{Tonry81}).
Since the first postulation almost 20 years ago of a huge nearby ``Great Attractor'' other than the
 Virgo-Cluster, (e.g. Shaya \cite{Shaya84}, 
Tammann \& Sandage \cite{Tamman85}, Aaronson et al. \cite{Aarons86} and \cite{Aarons89}), its 
approximate position, 
namely in direction to the Centaurus
super-cluster, has not changed. However, the distance difference between the GA and 
the Centaurus cluster has remained 
uncertain. A first robust value of
the GA-distance to the Local Group was derived by Lynden-Bell et al. (\cite{Lynden88}) 
using a projection of the Fundamental Plane (Djorgovski \& Davis \cite{Djorgo87}, 
Dressler et al. \cite{Dressl87}). 
They derived the GA to be located at a CMB radial velocity 
of 4350 $\pm$ 350 km/s. This corresponds to 62 Mpc for $H_0=$70 km/s/Mpc and is
significantly behind the Centaurus cluster. Later, Tonry et al. ({\it SBF II}) 
refined this measurement using 
SBF-distances, which resulted in a somewhat closer GA distance of 43 $\pm$ 3 Mpc. Still,
this was slightly behind the position of the Centaurus cluster, which they determined to be 
centered at 33 Mpc.\\
\subsection{The Centaurus cluster}
The proximity of the Centaurus cluster to the GA makes it an interesting subject 
for distance determination and to study possible effects of the GA's strong 
gravitational potential on the cluster's structure. If the Centaurus cluster was elongated
significantly in front of and behind the GA, an infall pattern would be expected,
which is an anti-correlation between redshift and distance caused by the GA's gravitational potential 
distorting the Hubble flow (e.g. Dressler \& Faber \cite{Dressl90}).\\ 
An 
additional feature makes the Centaurus cluster even more attractive -- but at the same time more complex --: 
in redshift space it consists of two well separated sub-clusters, namely the dominating 
component Cen30 at about 3000km/s and the 2-3 times smaller component Cen45 at about 
4500 km/s. 
In several studies this remarkable substructure has been investigated (e.g. Lucey 
et al. \cite{Lucey80} \& \cite{Lucey86}, Jerjen et al. \cite{Jerjen97}, Stein et al. 
\cite{Stein97}, Churazov et al. \cite{Churaz99}, Furusho et al. \cite{Furush01}), 
indicating that Cen45 is probably a subgroup falling into the main cluster Cen30. 
Lucey et al. (\cite{Lucey86}) suggest that Cen45 is located at about the same distance 
as Cen30, based mainly on a comparison of the cumulative luminosity distribution in 
both sub-clusters. Churazov et al. (\cite{Churaz99}) propose, based on {\it ASCA} X-ray temperature 
measurements, that the two subcomponents are merging. They suggest the existence
of a large scale filament along the line of sight towards Centaurus in order to explain
the discrepancy between the unusually high velocity dispersion of the Cen30 members and the X-ray 
temperature.
Furusho et 
al. (\cite{Furush01}) present more extended X-ray measurements and conclude that 
a major merger in Centaurus rather occurred several Gyrs ago.
Stein et al. (\cite{Stein97}) find that the morphological content 
of the two sub-clusters differs substantially. Cen30 is more dominated by early-type 
galaxies, while Cen45 contains more late-type galaxies and fewer dwarfs. This is 
consistent with Cen30 being the older, main cluster, and Cen45 the more active young 
infalling sub-cluster.\\
\subsection{Centaurus cluster galaxy distances with Surface Brightness Fluctuations}
A promising possibility to determine a precise Centaurus cluster distance and 
gain more insight into its spatial structure is by deriving galaxy distances using the 
Surface Brightness Fluctuations (SBF) method (Tonry \& Schneider \cite{Tonry88}).
The first published SBF-distances to Centaurus cluster galaxies were made by Dressler
(\cite{Dressl93}), who derived distances to four Cen30 and two Cen45 members. He obtained a 
distance modulus of about 32.1 mag for Cen30 and 32.2 mag for Cen45, yielding
high peculiar velocities of about 1400 km/s for the investigated galaxies.\\
Later, these measurements were refined and complemented by Tonry et al. (Tonry et al. \cite{Tonry97};
and Tonry et al. \cite{Tonry01}, in the following {\it SBF IV}) in the course of
their SBF survey, resulting in a somewhat greater distance for Centaurus.
They obtained distance moduli to 5 Cen30 
and 3 Cen45 members. The resulting mean distance moduli are 32.51 $\pm$ 0.11 mag for 
Cen30 and 32.80 $\pm$ 0.09 mag for Cen45, showing a distance 
difference between the two sub-clusters at 1.45 $\sigma$ significance. However, 
already in {\it SBF IV} and Blakeslee et al. (\cite{Blakes02}) it has been pointed out
that these results are subject to a selection effect biasing towards closer distances by up to
0.3 mag: the sensitivity of Tonry's survey
is reached at about the distance of the Centaurus cluster. This makes those galaxies whose
observational and statistical errors place them closer than the mean cluster distance more probable to be
included in their survey than those who fall behind the cluster for their errors. A
discussion of this will be given in Sect.~\ref{discussion} of this paper.\\\\
Deeper and more numerous SBF measurements than those of Tonry et al. are needed 
in order to reduce this 
selection effect and allow a less biased calculation of the Centaurus cluster distance.
\subsection{Aim of this paper}
To improve the distance precision to the entire Centaurus cluster and its subcomponents, 
we present in this paper new SBF distance measurements to 15 early type Centaurus 
galaxies -- 3 giants and 12 dwarfs. The data originate from deep VLT 
FORS1 exposures in the $I$-band of six different $7\times7$' fields in the central 
Centaurus cluster. Of the 15 galaxies, 11 have measured radial velocities. 8 of them 
belong to Cen30 and 3 to Cen45. The two major giant elliptical galaxies of Cen30 and 
Cen45, namely NGC 4696 and NGC 4709, constitute the overlap between Tonry's and 
our dataset.\\
SBF have been measured for only small 
samples of dwarf galaxies, yet (e.g. Bothun \cite{Bothun91},
Jerjen et al. \cite{Jerjen98}, \cite{Jerjen00} \cite{Jerjen01} and Jerjen \cite{Jerjen03}). 
Recently, Mieske et al. (\cite{Mieske03}, in the following {\it MieskeI}) have 
presented SBF simulations 
to test the potential of the SBF-Method to measure distances to dwarf galaxies.
To our knowledge, the sample presented in this paper is the largest homogeneous sample 
of dwarf galaxies with SBF-distances up to now.\\
The paper is structured as follows: Section \ref{derive} explains how 
the absolute fluctuation magnitude $\overline{M}_I$ is derived from $(V-I)$. 
In section \ref{datareduction}, the data and their 
reduction are described. Section \ref{results} shows the results of the SBF 
measurements. They are discussed in section \ref{discussion}. We finish this paper 
with the conclusions in section \ref{conclusions}.\\
\section{Deriving $\overline{M}_I$ from $(V-I)$}
\label{derive}
\begin{figure}
\begin{center}
\psfig{figure=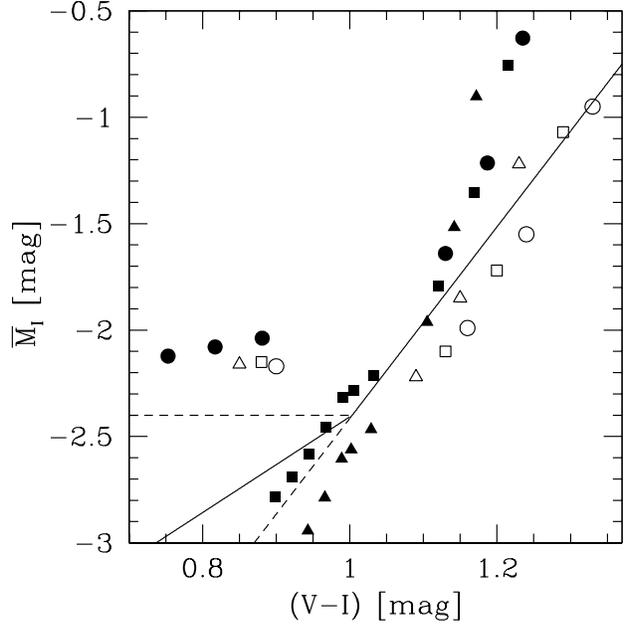,width=8.6cm}
\end{center}
\caption[]{\label{visbftheo}Theoretical values of $\overline{M}_I$ plotted vs. $(V-I)$ 
for a range of metallicities (-1.7 to 0 dex) and ages (8, 12 and 17 Gyr), taken from 
Liu et al. (\cite{Liu00}) (open symbols) and Worthey (\cite{Worthe94}) (filled symbols). 
The different symbols represent different ages as follows: triangle 8 Gyrs, 
square 12 Gyrs, circle 17 Gyrs. For the 17 Gyr Worthey-models, the lowest metallicity
is -1.3 dex. Metallicity increases 
towards redder $(V-I)$ and fainter $\overline{M}_I$. 
The solid line represents the $\overline{M}_I$--$(V-I)$ relation
adopted in this paper. The dashed lines
for $(V-I)\le1.0$ indicate the constant $\overline{M}_I=-2.40$ mag and equation~(\ref{sbfrel})'s
continuation (see text for further details).}
\end{figure}

The observable derived from SBF measurements is the apparent fluctuation magnitude 
$\overline{m}$, equivalent to the luminosity weighted mean apparent luminosity of 
the stellar population. To estimate the distance of a galaxy with the SBF-Method,
the absolute fluctuation magnitude 
$\overline{M}$ must be derived from a distance independent observable.\\
Tonry et al. (\cite{Tonry97}, in the following {\it SBF I};
and {\it SBF IV}) 
have established an empirical calibration between colour $(V-I)$ and the absolute 
fluctuation magnitude $\overline{M}_{\rm I}$:
\begin{equation}
\overline{M}_{\rm I}=-1.74 + 4.5 \times ((V-I) - 1.15)\;{\rm mag}
\label{sbfrel}
\end{equation}
According to Tonry et al. ({\it SBF IV}), this equation is valid only for galaxies with 
$(V-I)>1.0$. As the galaxies investigated in this paper span a somewhat 
larger range of colour, $0.84<(V-I)<1.35$ mag (see table~\ref{galprob}), it must be investigated
whether for the three galaxies in our sample with $(V-I)<1.0$ mag, equation~(\ref{sbfrel}) 
can be applied.\\
There is currently no published data available for $I$-band SBF measurements for $(V-I)<1.0$ mag. 
However, Jerjen et al. (\cite{Jerjen98}, \cite{Jerjen00}) conclude from $R$-band
SBF measurements of nearby dEs that for blue $(B-R)$ the relation between
$\overline{M}_{\rm R}$ and $(B-R)$ bifurcates into a steep one continuing the relation
found for red dEs and a shallower one, giving fainter $\overline{M}_{\rm R}$ at a given
$(B-R)$.\\
To look into this in more detail, in Fig.~\ref{visbftheo} theoretical values of $\overline{M}_I$ vs. 
$(V-I)$ are plotted for a range of metallicities 
and ages typical for early type galaxies, using models of Worthey (\cite{Worthe94}) and 
Liu et al. (\cite{Liu00}). 
For red colours, both models trace equation~(\ref{sbfrel}) well, with the Worthey-models 
being more
deviant from the empirical calibration than the Liu-models. In the blue range, the Worthey-models 
for 8 and 12 Gyrs trace equation~(\ref{sbfrel}), while the 17 Gyr Worthey-models and the metal-poor 
Liu-models predict $\overline{M}_I$ substantially fainter 
than according to equation~(\ref{sbfrel}).
This is in agreement with Jerjen's findings and shows that a correction of equation~(\ref{sbfrel})
towards fainter $\overline{M}_I$ should be applied for blue colours. 
To this end, we follow the same line of argument made 
in {\it MieskeI}, such that for $(V-I)<1.0$ we adopt the average of $\overline{M}_I$
according to equation~(\ref{sbfrel}) and a constant value of $\overline{M}_I=-2.4$ mag, which is
$\overline{M}_I$ at $(V-I)=1.0$. This yields the following equation holding for $(V-I)<1.0$:
\begin{equation}
\overline{M}_{\rm I}=-2.07+2.25 \times ((V-I) - 1.15)\;{\rm mag}
\label{sbfrel3}
\end{equation}
In Fig.~\ref{visbftheo}, this is indicated graphically. For $(V-I)\ge1.0$, equation~(\ref{sbfrel}) is adopted.\\
Note that the adjustment for $(V-I)<1.0$
should serve to decrease possible
systematic biases occurring when applying an unchanged equation~(\ref{sbfrel}), but it
should not be considered especially accurate. The lack of observational data in this colour regime
and the discrepancy between different model grids forces us to adopt a ``best guess''. 
After all it is worth remarking that the changes discussed 
only apply to 3 of the 15 sample galaxies and are of the order of 0.3 mag or smaller.\\
It is necessary to define the intrinsic uncertainty of the two above equations,
also referred to as cosmic scatter. The cosmic scatter of equation~(\ref{sbfrel}) has already been
determined by Tonry et al. in {\it SBF IV} to be between 0.05 and 0.10 mag. They base this finding
on the scatter of their SBF distance measurements for galaxies with known distances. Tonry et al. 
investigated only giant galaxies. As we investigate giants and dwarfs, we do not expect a 
smaller spread in stellar contents than if investigating only giants. We therefore adopt the upper limit
of 0.10 mag as the cosmic scatter for $(V-I)\ge1.0$. For $(V-I)\le1.0$, we adopt the quadratic sum of 
the former 0.10 mag scatter and the magnitude difference between equation~(\ref{sbfrel3}) and equation~(\ref{sbfrel}). This amounts up to 0.37 mag for the bluest of our investigated galaxies ($(V-I)$=0.84).\\ 
\section{The data}
\label{datareduction}
\begin{figure}
\begin{center}
\psfig{figure=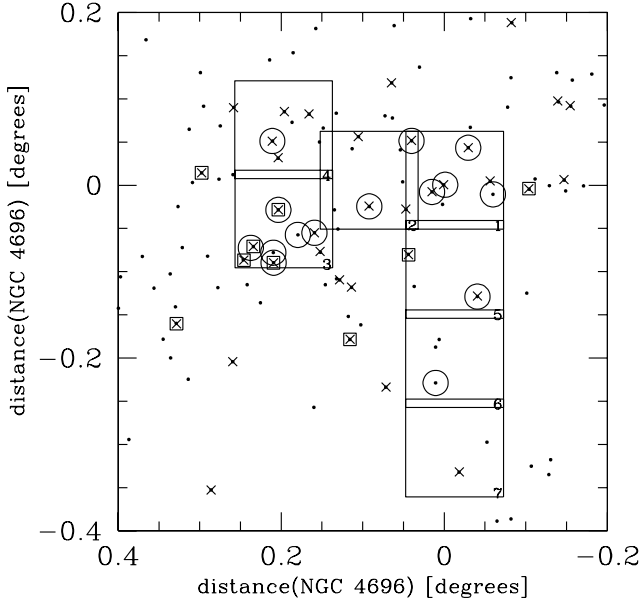,width=8.6cm}
\end{center}
\caption[]{\label{mapofcen}Map of the central Centaurus cluster, with distances relative 
to the main galaxy of Cen30, NGC 4696. East is left, North is up. 
Large squares are the observed VLT fields, the 
field number is indicated in the lower right corner of each square. Dots represent the 
galaxies listed in Jerjen et al.'s (\cite{Jerjen97}) Centaurus Cluster Catalog (CCC) as 
probable and likely cluster members. Dots marked with crosses are galaxies for which Stein 
et al. (\cite{Stein97}) have obtained radial velocity. Small squares indicate galaxies 
belonging to the sub-cluster Cen45. Galaxies marked with large circles are the ones for 
which we present new SBF measurements in this paper.}
\end{figure}

The data for this publication have been obtained in service mode at the Very Large 
Telescope (VLT) of the European Southern Observatory,
Chile (Observing Programme 67.A--0358), using UT 1 with the instrument FORS1 in imaging 
mode. Seven $7\times7$' fields in the central Centaurus cluster have been observed in 
Johnson $V$ and $I$ pass-bands. The seeing ranged between 0.4 and 0.6$''$. The total 
integration time was 1500 seconds for the $V$ exposures, divided up into 4 dithered 
single exposures, and 3000 seconds for the $I$ exposures, divided up into 9 dithered 
single exposures. Fig.~\ref{mapofcen} shows a map of the central Centaurus cluster 
with the observed fields and cluster galaxies indicated. Table~\ref{obs} gives the 
coordinates of the observed fields. Table~\ref{galprob} gives the photometric 
properties and coordinates of the 15 investigated cluster galaxies. They span a magnitude
range of $19.6>V>11.5$ mag, corresponding to approximately $-21.5<M_{\rm V}<-13.5$.\\
In the 7 fields, there are located 14 additional galaxies 
cataloged as early-types in the CCC. These could not be investigated for the following 
reasons: 5 galaxies showed pronounced spiral features on
our high resolution images, revealing that they are probably late-type background galaxies rather
than early-type cluster members; 6 galaxies were too faint to detect a significant SBF signal; 2
galaxies showed pronounced boxy residuals after subtracting an elliptical light model, with the
boxy features having a scale size of only a few times that of the seeing; 1 galaxy was too close
to the halo of a bright saturated star to obtain a reliable SBF signal.
\begin{table*}
\begin{center}
\begin{tabular}{lllllrlll}
Field & RA [2000] & Dec [2000] & ZP$_{\rm I}$ & ZP$_{\rm V}$ & CT$_{\rm I}$ & CT$_{\rm V}$ & k$_{\rm I}$ & k$_{\rm V}$\\\hline 
1 & 12:48:45.0 & -41:18:20  & 26.582  & 27.472  & 0.00    & 0.08  & 0.093  & 0.145  \\
2 & 12:49:18.5 & -41:18:20  & 26.60   & 27.446  & 0.00    & 0.08  & 0.093  & 0.145  \\
3 & 12:49:52.0 & -41:21:02  & 26.672  & 27.514  & -0.088  & 0.015 & 0.093  & 0.145  \\
4 & 12:49:52.0 & -41:14:50  & 26.588  & 27.446  & 0.00  & 0.08 & 0.093  & 0.145  \\
5 & 12:48:45.0 & -41:24:32  & 26.548  & 27.446  & 0.00  & 0.08 & 0.093  & 0.145  \\
6 & 12:48:45.0 & -41:30:44  & 26.548  & 27.446  & 0.00  & 0.08 & 0.093  & 0.145  \\
7 & 12:48:45.0 & -41:36:56  & 26.582  & 27.472  & 0.00  & 0.08 & 0.093  & 0.145  \\
\end{tabular}
\end{center}
\caption[]{\label{obs}Central coordinates and photometric calibration coefficients 
for the 7 VLT FORS1 fields as indicated in Fig.~\ref{mapofcen}.}
\end{table*}
\begin{table*}
\begin{center}
\begin{tabular}{llllllll}
CCC-Nr.$^*$ & Field & RA$^*$ [2000] & Dec$^*$ [2000] & V$_0$$^{**}$ [mag] & (V-I)$_0$$^{**}$ [mag] & $v_{rad}$ [km/s] & Type$^*$ \\\hline 
52      & 1     &  12:45:44.3 & -41:02:58 & 17.86       & 1.09 & ---      & dE,N \\
61      & 1     &  12:48:39.7 & -41:16:05 & 16.26       & 1.14 & 2910     & dE,N \\
65 (NGC 4696)   & 1     &  12:48:49.0 & -41:18:39 & 11.50        & 1.24 & 2985     & E4,S03(4)  \\
70      & 1     &  12:48:53.9 & -41:19:09 & 16.69       & 1.24 & 2317     & cdE (E0 in CCC)   \\
75      & 1     &  12:49:01.9 & -41:15:36 & 17.31       & 1.12 & 1958     & dE,N \\
89      & 2     &  12:49:18.2 & -41:20:07 & 15.43       & 1.15 & 3104     & E1   \\
111     & 3     &  12:49:40.0 & -41:21:59 & 15.86       & 1.01 & 2880     & dE,N \\
115     & 3     &  12:49:46.5 & -41:22:08 & 18.15       & 0.99 & ---     & dE   \\
121     & 3     &  12:49:54.2 & -41:20:24 & 17.36       & 1.07 & 4739     & dE (Im in CCC)   \\
123     & 3     &  12:49:56.1 & -41:24:04 & 17.35       & 1.03 & 4661     & dS0  \\
124     & 3     &  12:49:56.2 & -41:23:22 & 19.09       & 0.84 & ---     & dE   \\
130 (NGC 4709)    & 3     &  12:50:04.0 & -41:22:57 & 12.5       & 1.35 & 4650     & E3 \\
125     & 4     &  12:49:56.4 & -41:15:37 & 16.06       & 1.08 & 2880     & dE,N \\
58      & 5     &  12:48:36.1 & -41:26:25 & 16.78       & 1.02 & 3304     & dE   \\
68      & 6     &  12:48:52.5 & -41:32:25 & 19.63       & 0.93 & ---     & dE   \\
\end{tabular}
\end{center}
\caption[]{\label{galprob}Coordinates and photometric properties of the investigated 
galaxies. The galaxies are ordered by field-number, and within the same field by 
right ascension. The field number refers to the fields indicated in Fig.~\ref{mapofcen}. 
The radial velocities are taken from Stein et al.'s (\cite{Stein97}) catalog of radial
 velocities for the Centaurus cluster. \\ 
$^*$As in the Centaurus cluster Catalog 
(CCC, Jerjen et al. \cite{Jerjen97}). Note that galaxy CCC 121 is cataloged in 
the CCC as being of type Im.
Based on our high resolution photometry, we cannot confirm this morphological type
but rather find it has a normal, smooth dE-like morphology. Galaxy CCC 70 is cataloged
as type E0 due to its relatively high surface brightness. However, it has $M_V\simeq-16.5$ mag,
placing it into the dwarf galaxy regime. We therefore adopt the type compact dE (cdE) for
CCC 70. $^{**}$Based on this paper.}
\end{table*}
\subsection{Data reduction before SBF measurement}
The pipeline reduced images still showed large scale sky count variations of the order 
of $\pm 3\%$. To partially compensate for that, a master flat field was constructed by 
combining all single exposures from all fields, disregarding contribution 
from astronomical objects to the final master flat using a sigma-clipping-rejection.
After division by the smoothed 
master flat field, the large scale variations were reduced to $\pm 1$ \%.\\
The observational zero points were obtained separately for each night.
 For all fields except for field 3, which 
was taken two months after the rest of the images, the colour terms were identical to 
within their errors. For all fields, the extinction coefficients were identical to 
within their errors. The accuracy of the derived zero points was of the order of 1\%. 
Table~\ref{obs} gives the calibration 
coefficients and central coordinates of all 7 Centaurus fields.\\
For each field and passband, the single exposures were brought into a common coordinate 
system by applying integer pixel shift corrections between the single dithered frames. 
For SBF measurements, only integer pixel shifts are suitable, as otherwise correlated 
noise would be introduced. Cosmic rays were removed from the single frames using the 
IRAF task COSMICRAYS. The registered cleaned single frames were averaged using 
an average sigma clipping algorithm.\\
For each investigated galaxy, the local background level was 
determined in both pass-bands via a curve of growth analysis, yielding the total apparent 
magnitudes in $V$ and $I$, a surface brightness profile 
and a colour map. To correct for galactic reddening and 
absorption, we used the values from Schlegel et al. (\cite{Schleg98}), who give $A_I=0.221$ and 
$E(V-I)=0.157$ for the coordinates of the Centaurus cluster.\\
\subsection{SBF measurement}
\begin{figure*}[ht!]
\begin{center}
\epsfig{figure=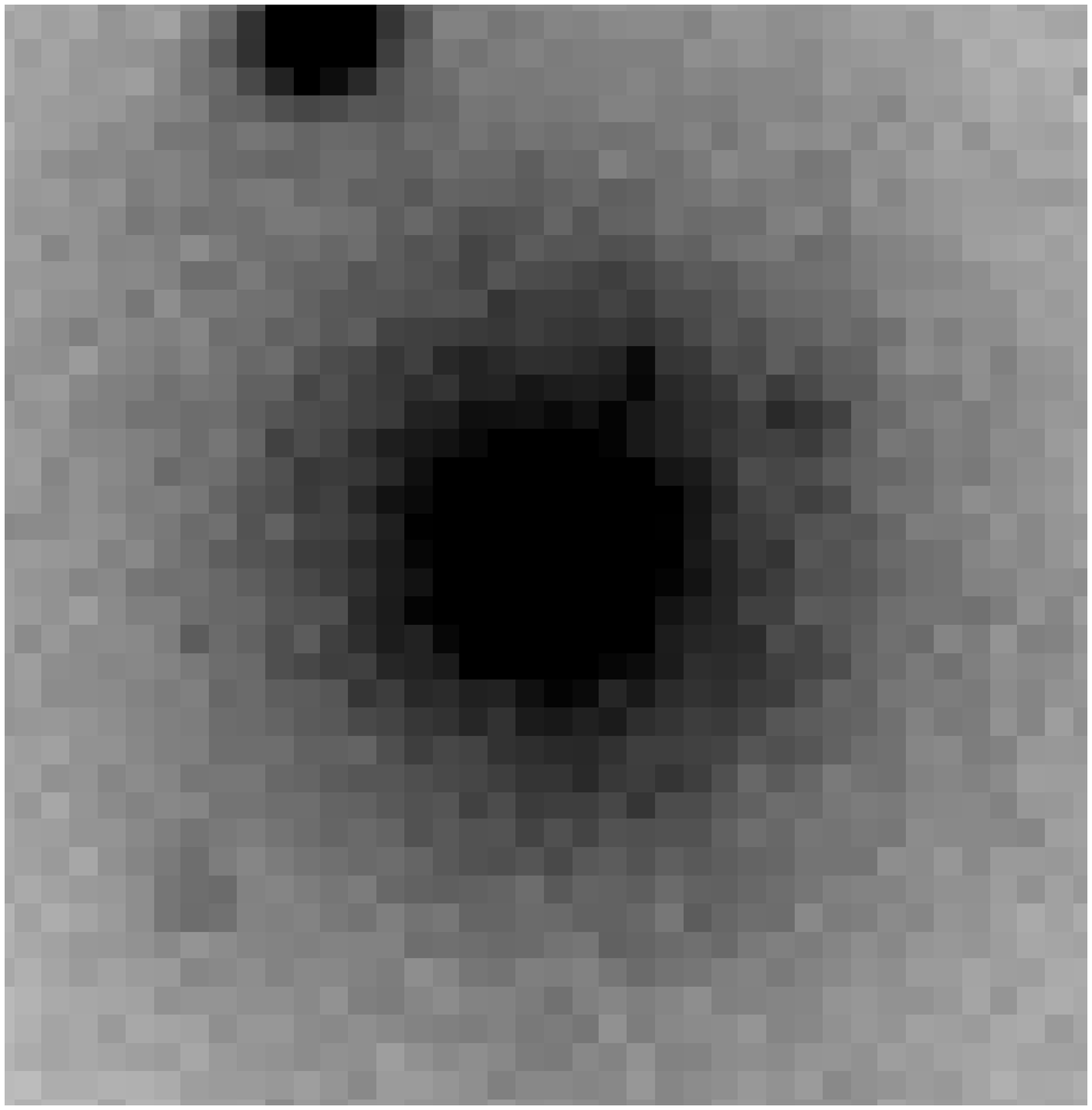,height=3.6cm,width=3.8cm}\hspace{0.2cm}
\epsfig{figure=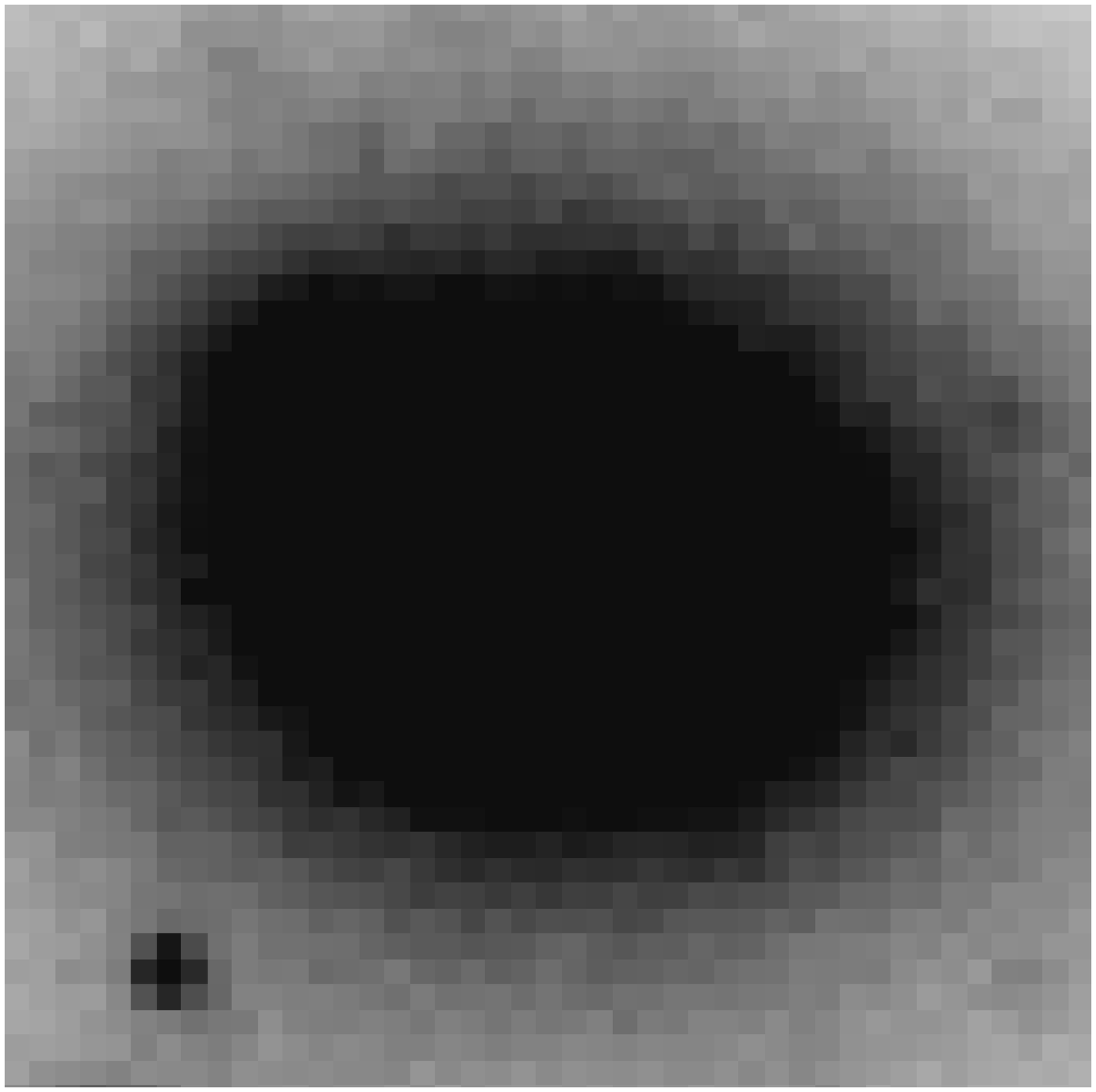,height=3.6cm,width=3.8cm}\hspace{0.2cm}
\epsfig{figure=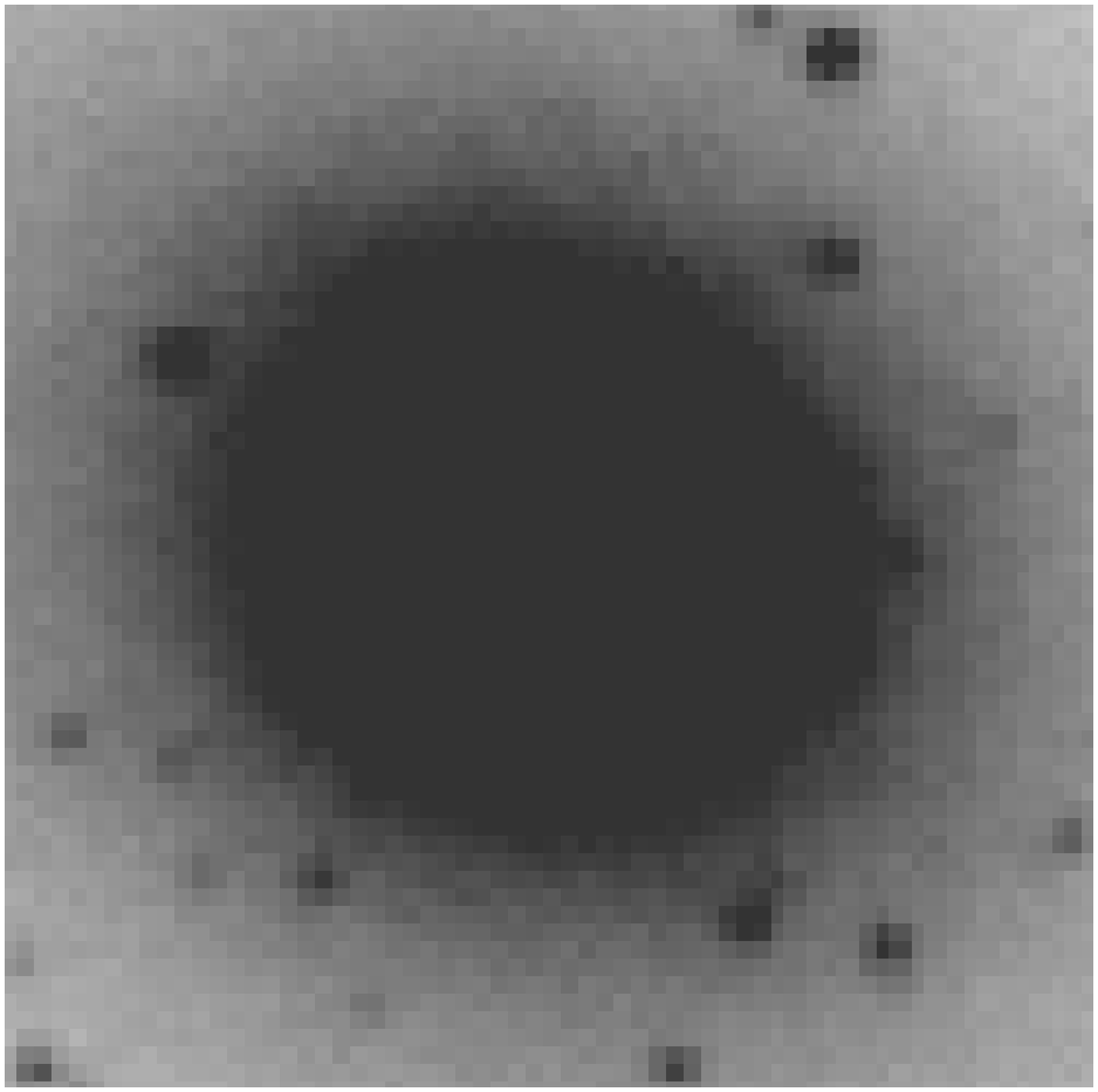,height=3.6cm,width=3.8cm}\hspace{0.2cm}
\epsfig{figure=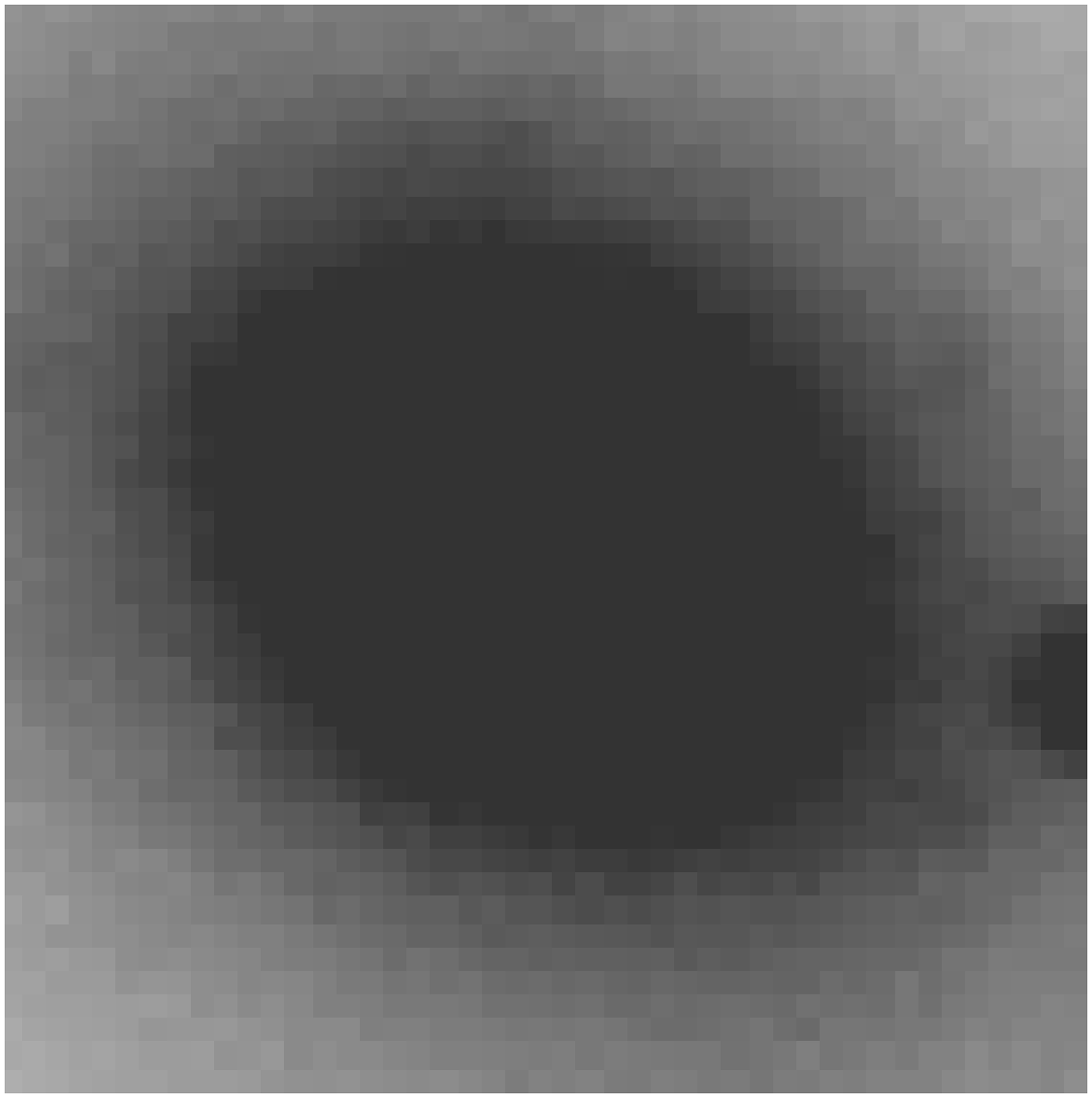,height=3.6cm,width=3.8cm}\\
\epsfig{figure=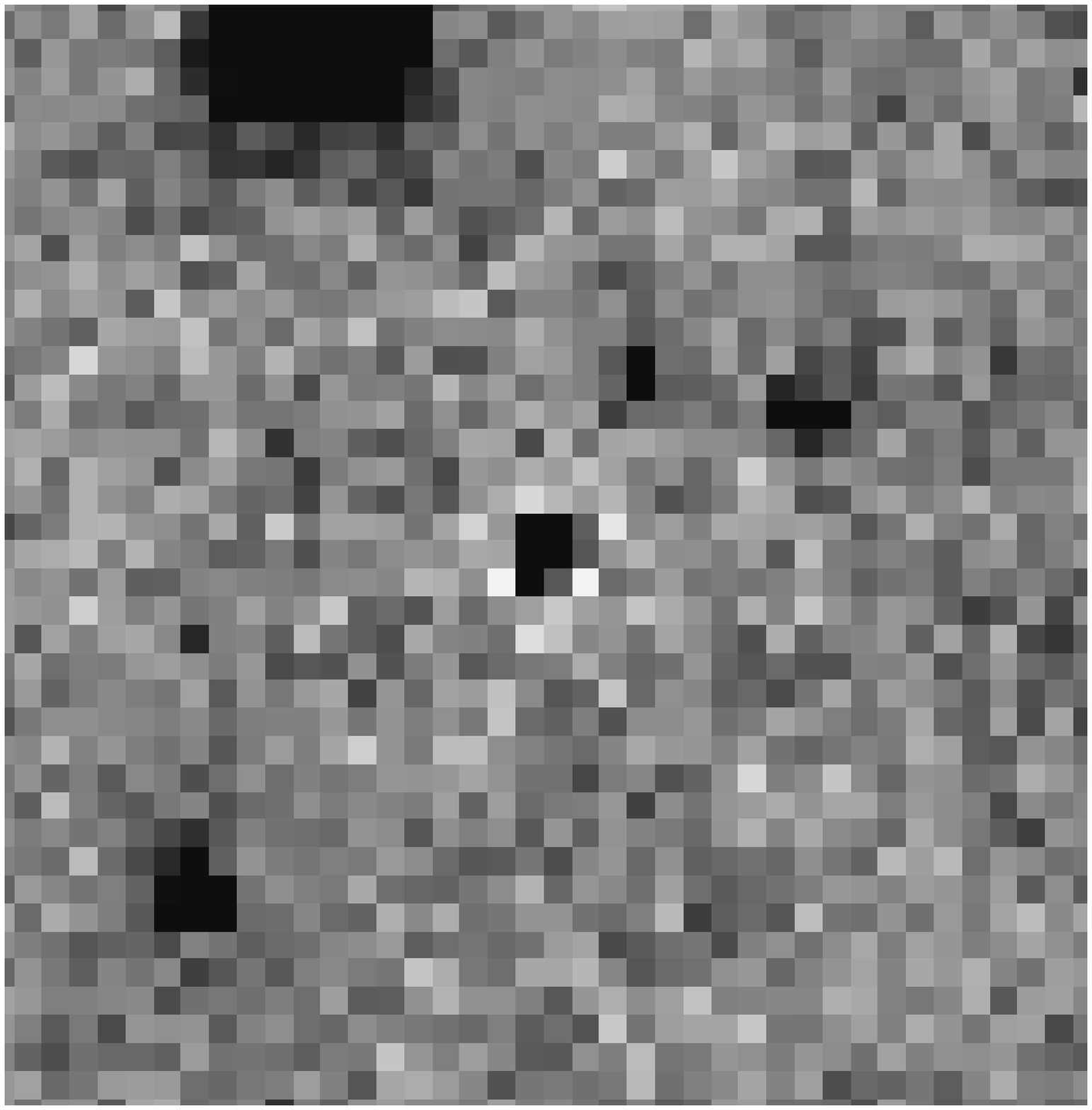,height=3.6cm,width=3.8cm}\hspace{0.2cm}
\epsfig{figure=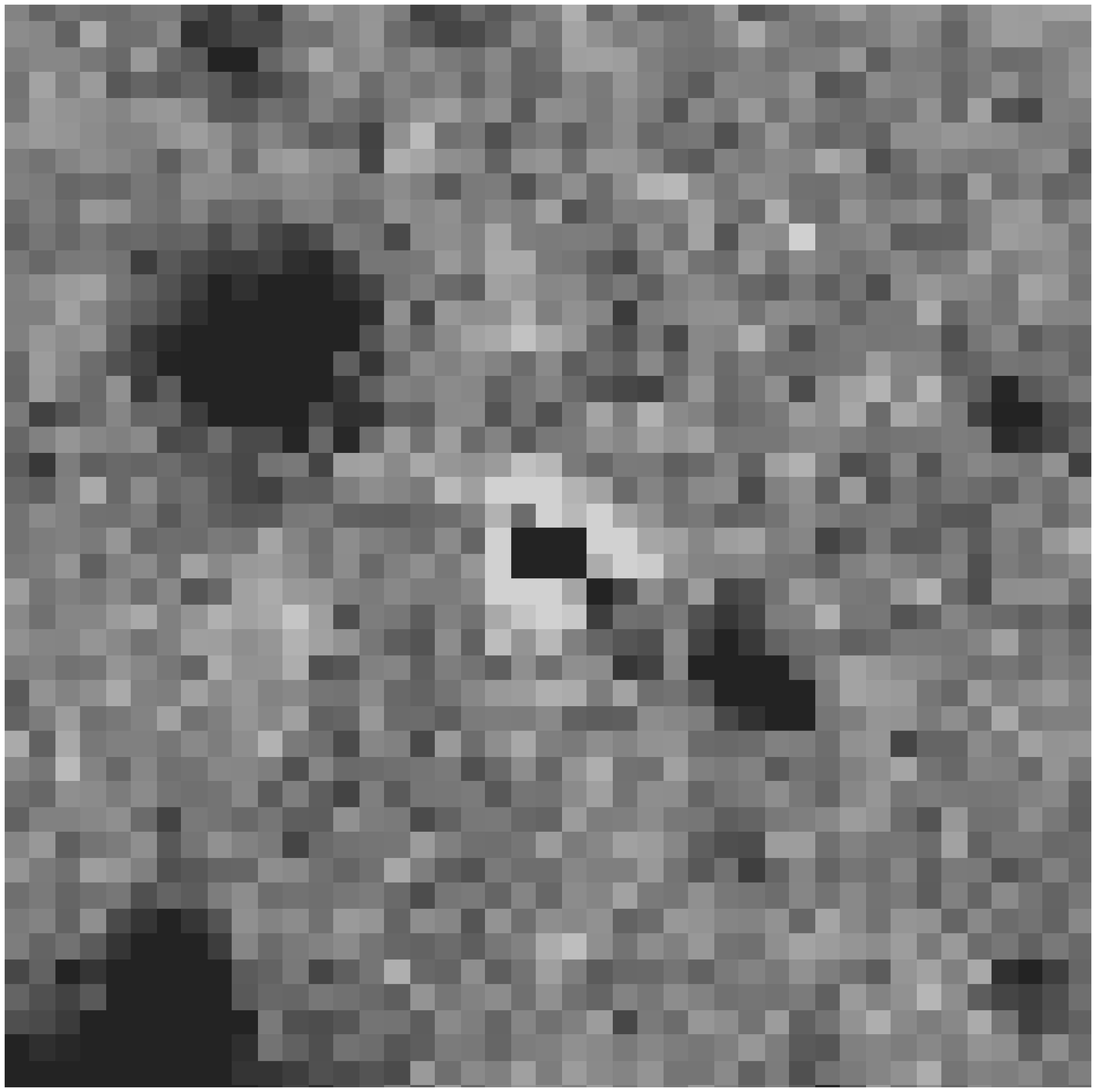,height=3.58cm,width=3.8cm}\hspace{0.2cm}
\epsfig{figure=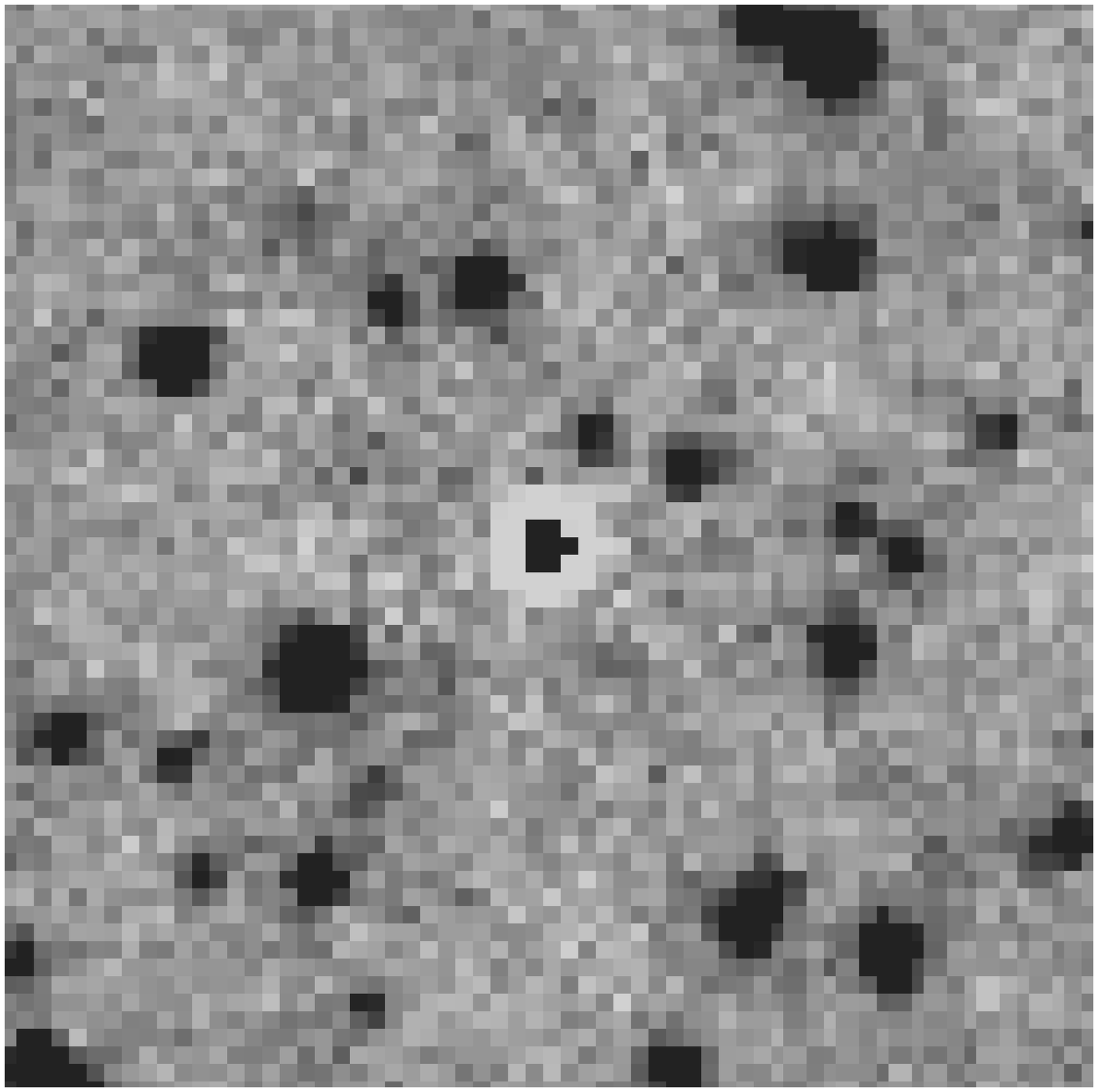,height=3.6cm,width=3.8cm}\hspace{0.2cm}
\epsfig{figure=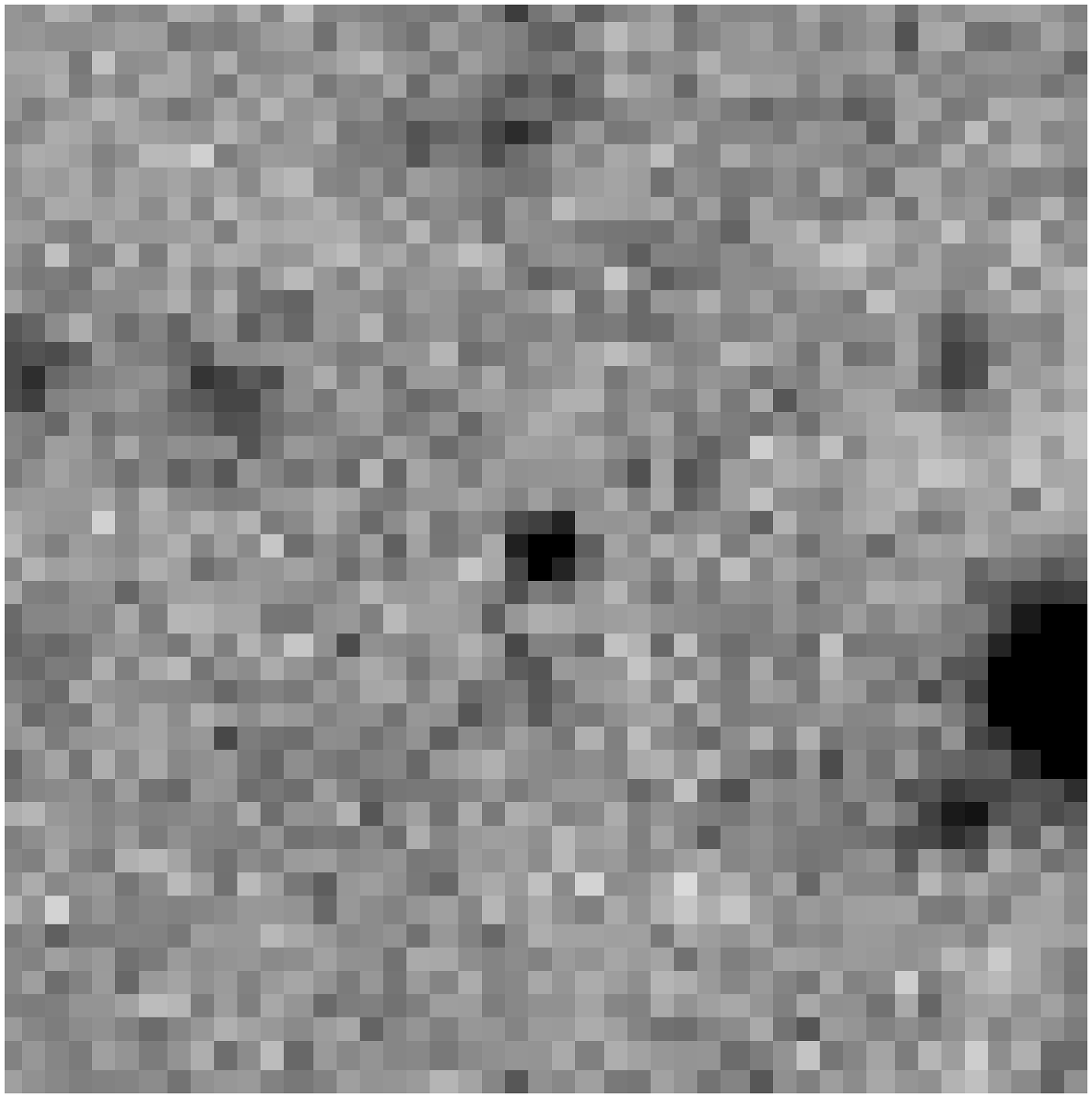,height=3.6cm,width=3.8cm}\\
\epsfig{figure=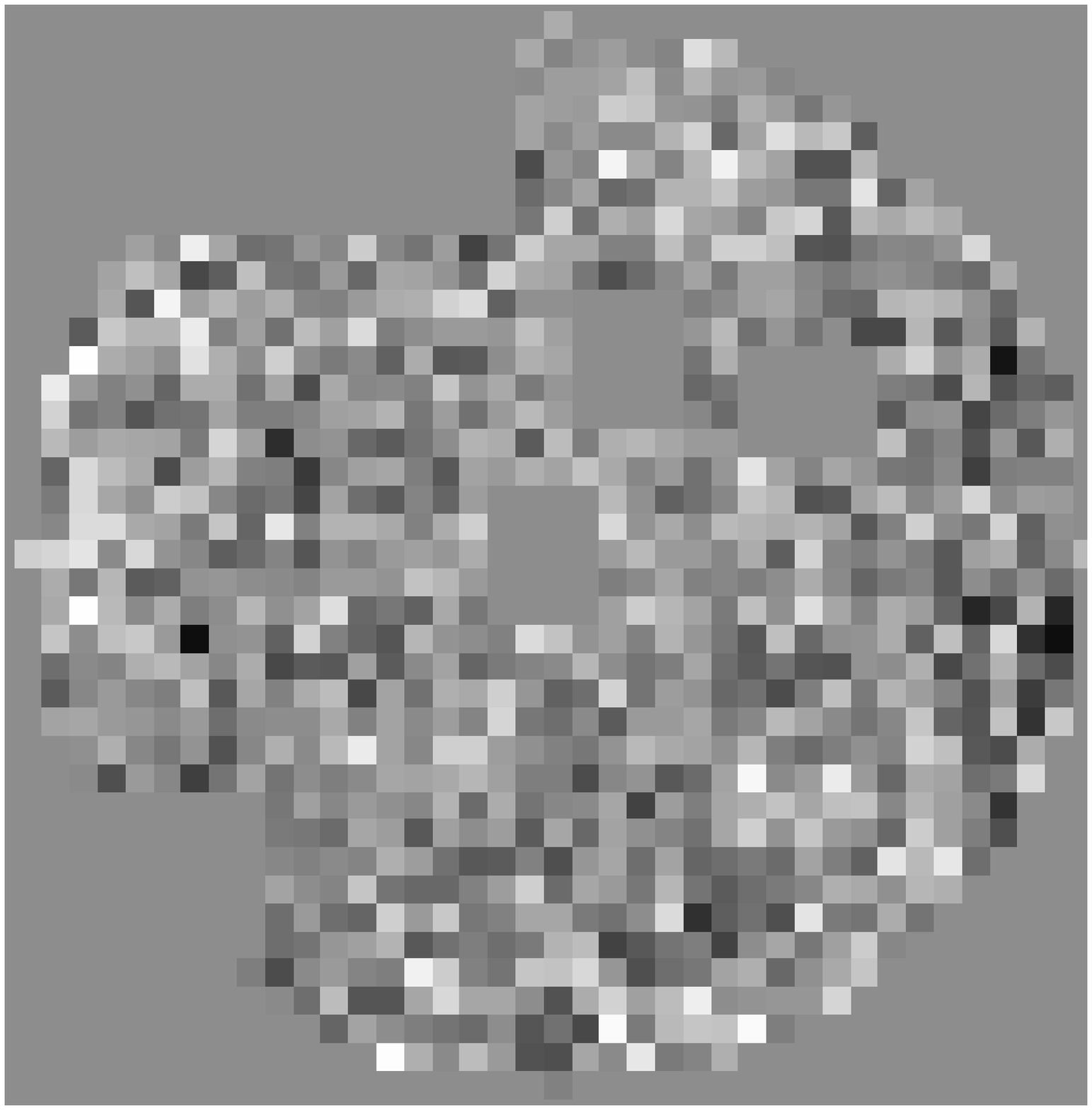,height=3.6cm,width=3.8cm}\hspace{0.2cm}
\epsfig{figure=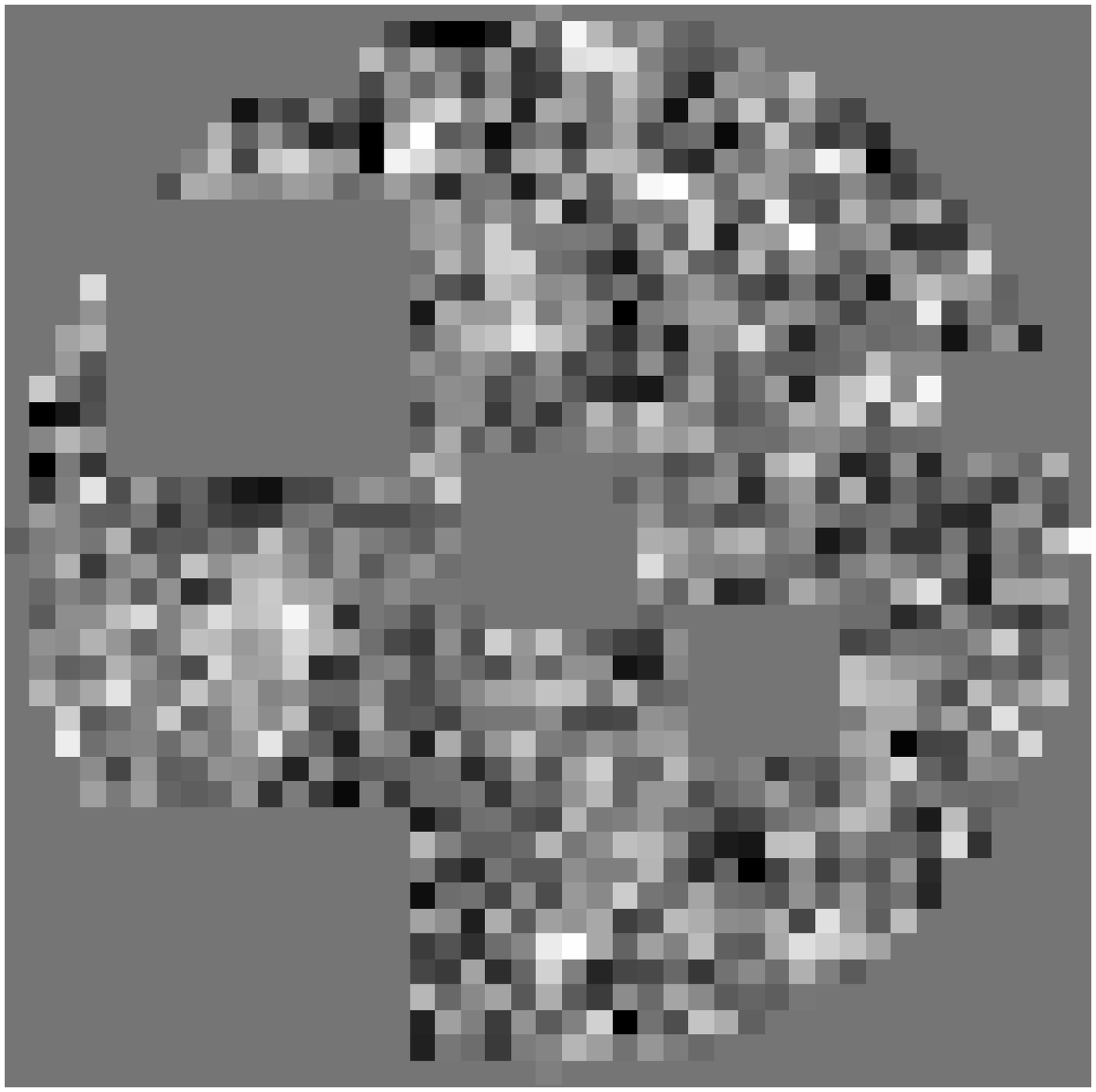,height=3.6cm,width=3.8cm}\hspace{0.2cm}
\epsfig{figure=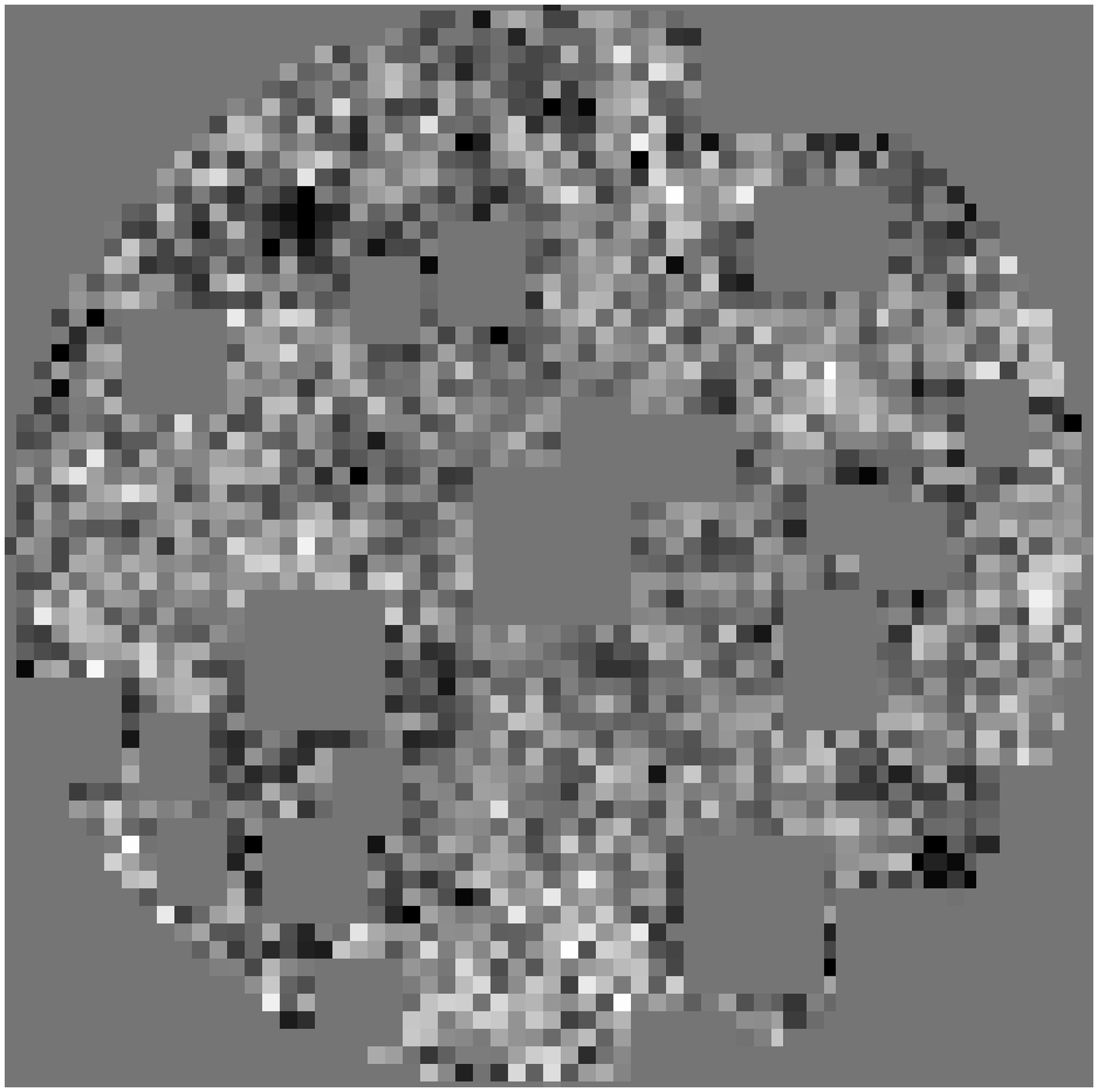,height=3.6cm,width=3.8cm}\hspace{0.2cm}
\epsfig{figure=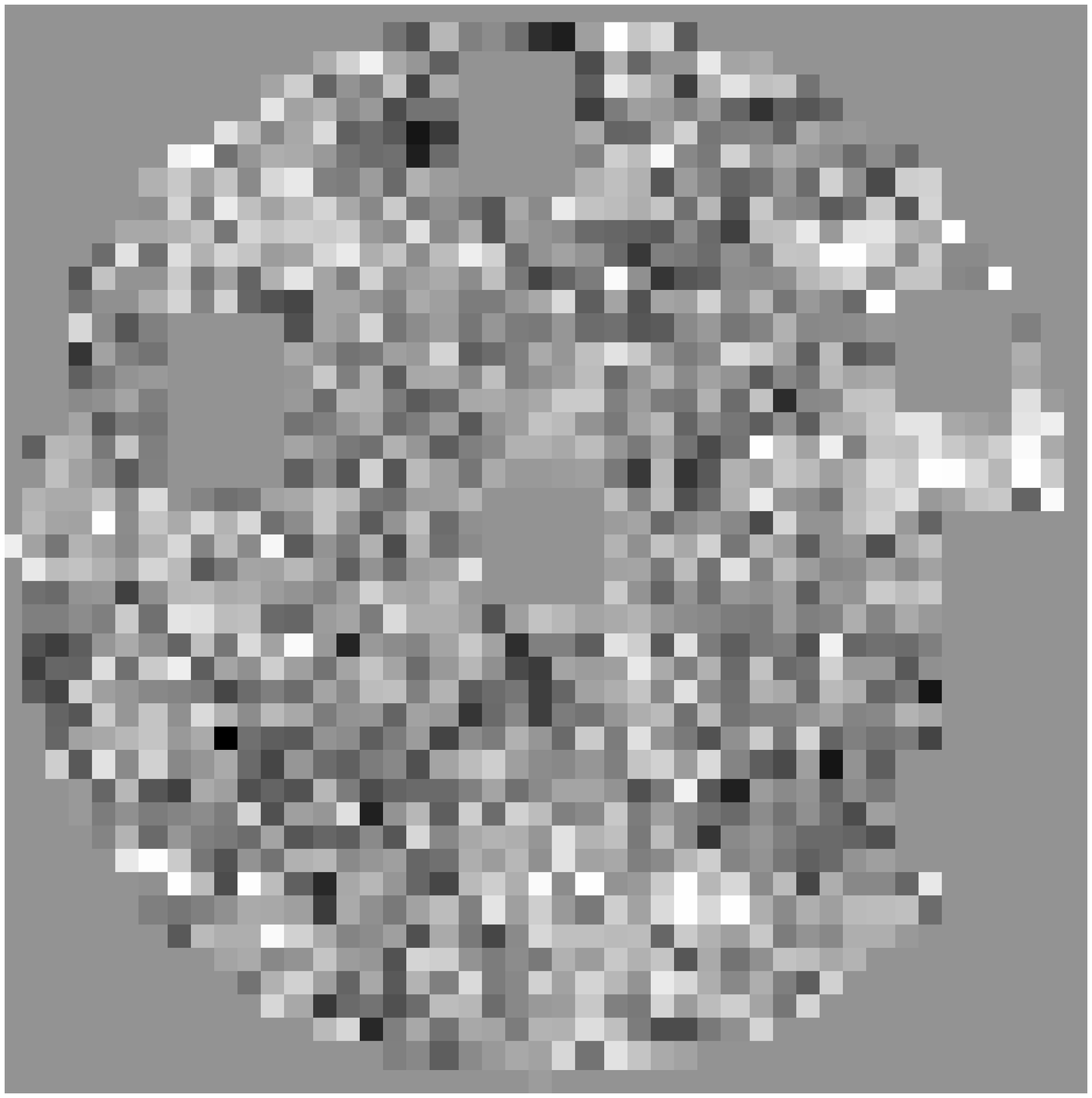,height=3.6cm,width=3.8cm}\\
\epsfig{figure=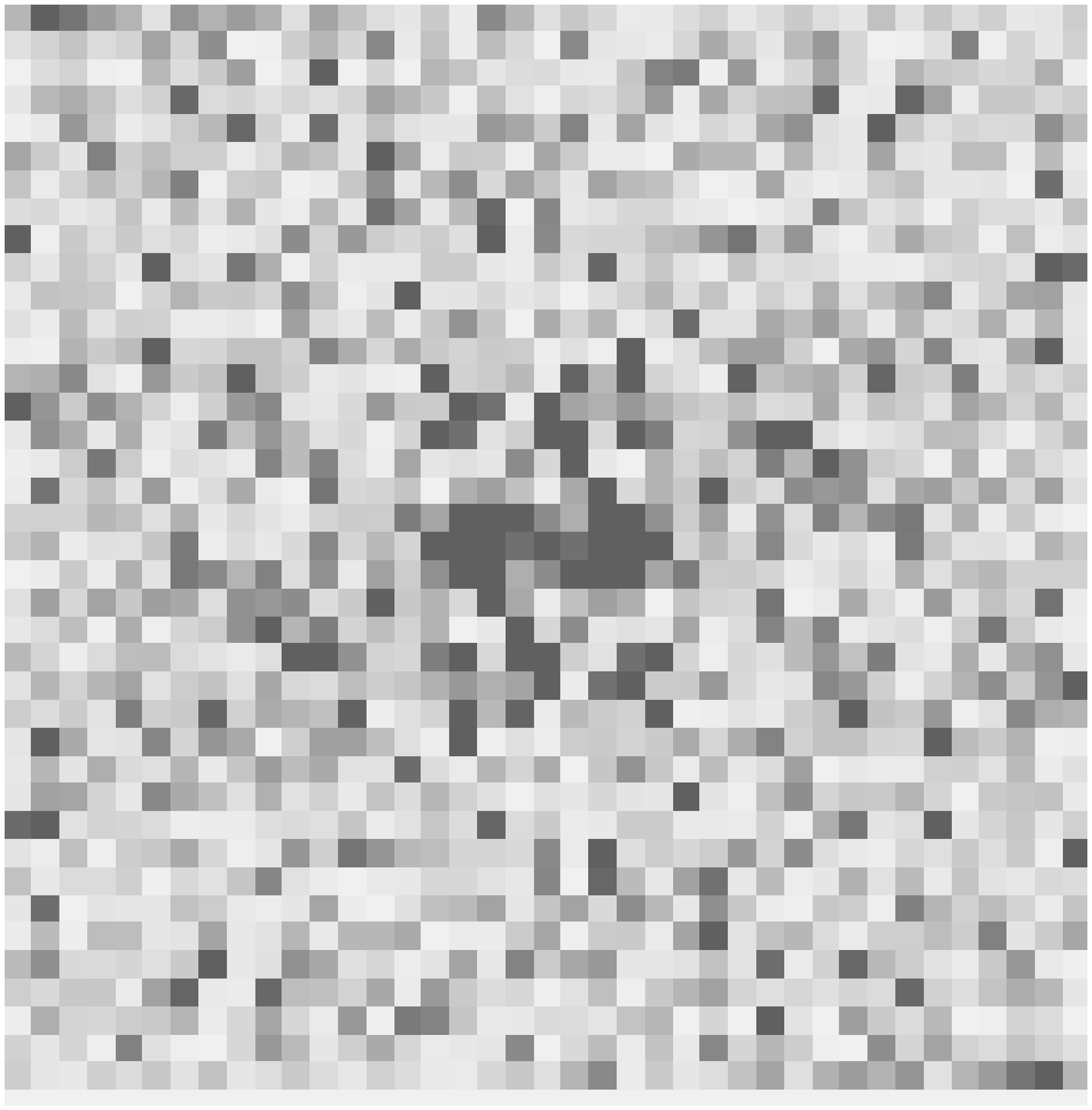,height=3.6cm,width=3.8cm}\hspace{0.2cm}
\epsfig{figure=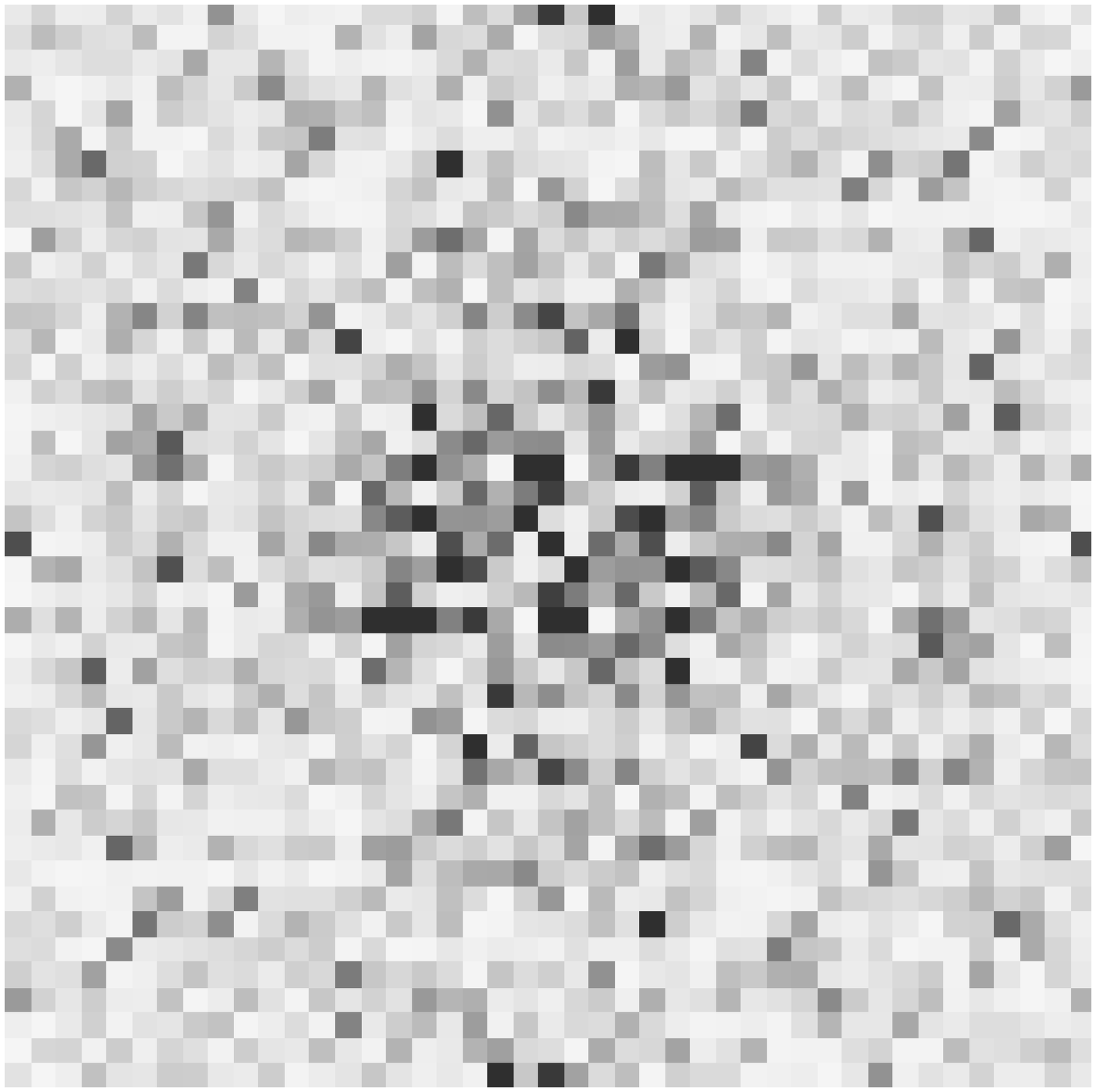,height=3.6cm,width=3.8cm}\hspace{0.2cm}
\epsfig{figure=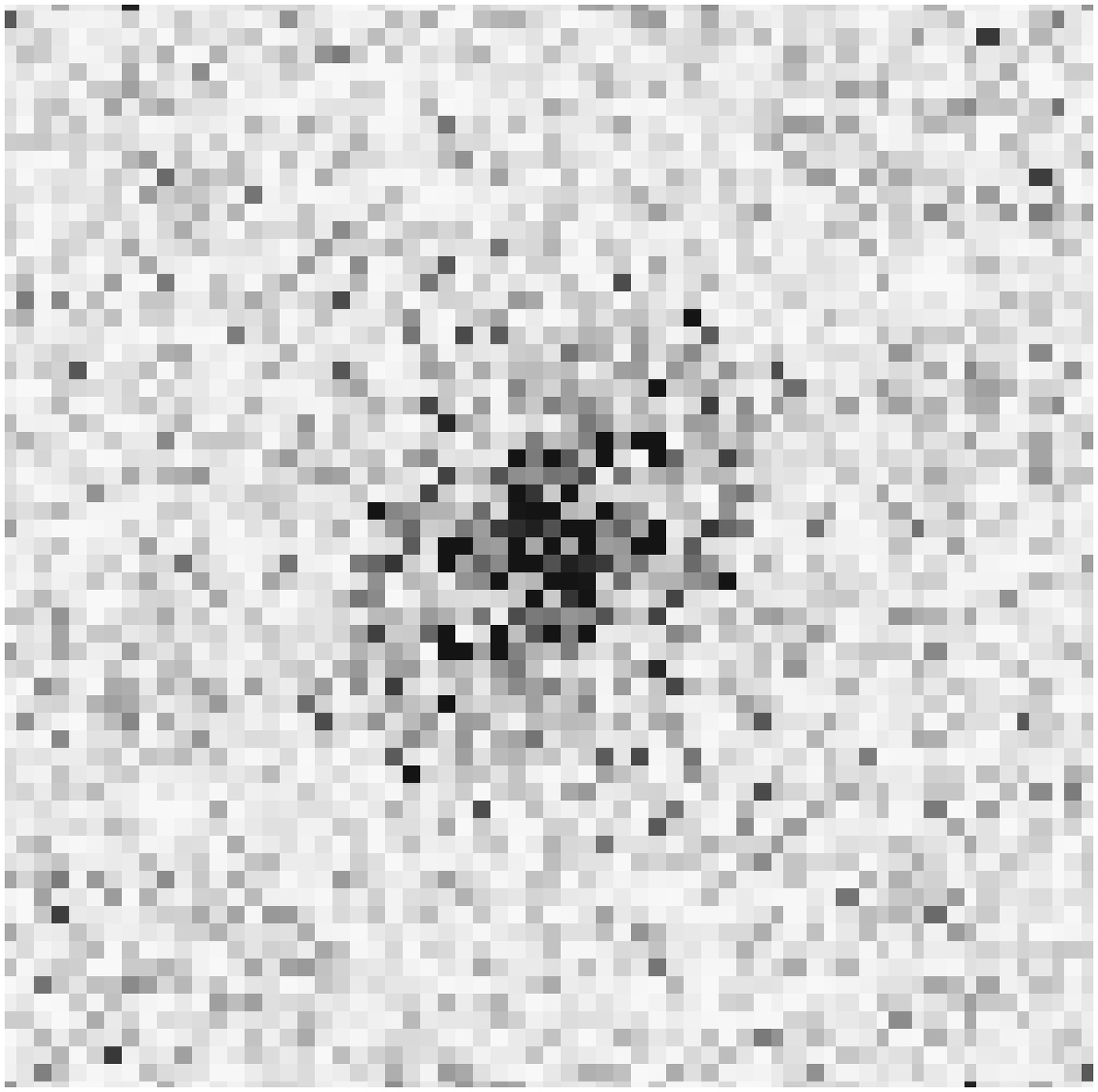,height=3.6cm,width=3.8cm}\hspace{0.2cm}
\epsfig{figure=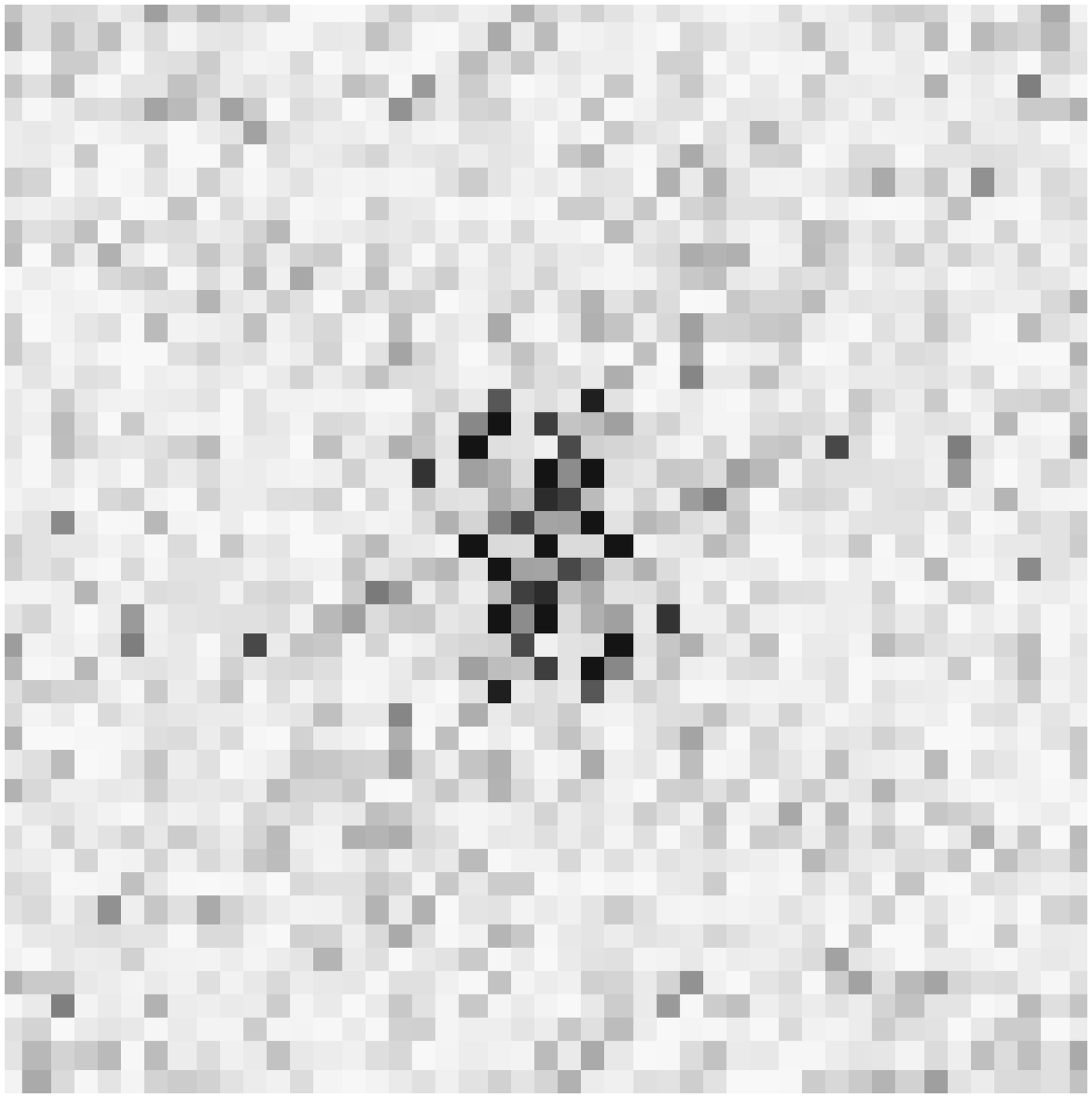,height=3.6cm,width=3.8cm}\\
\epsfig{figure=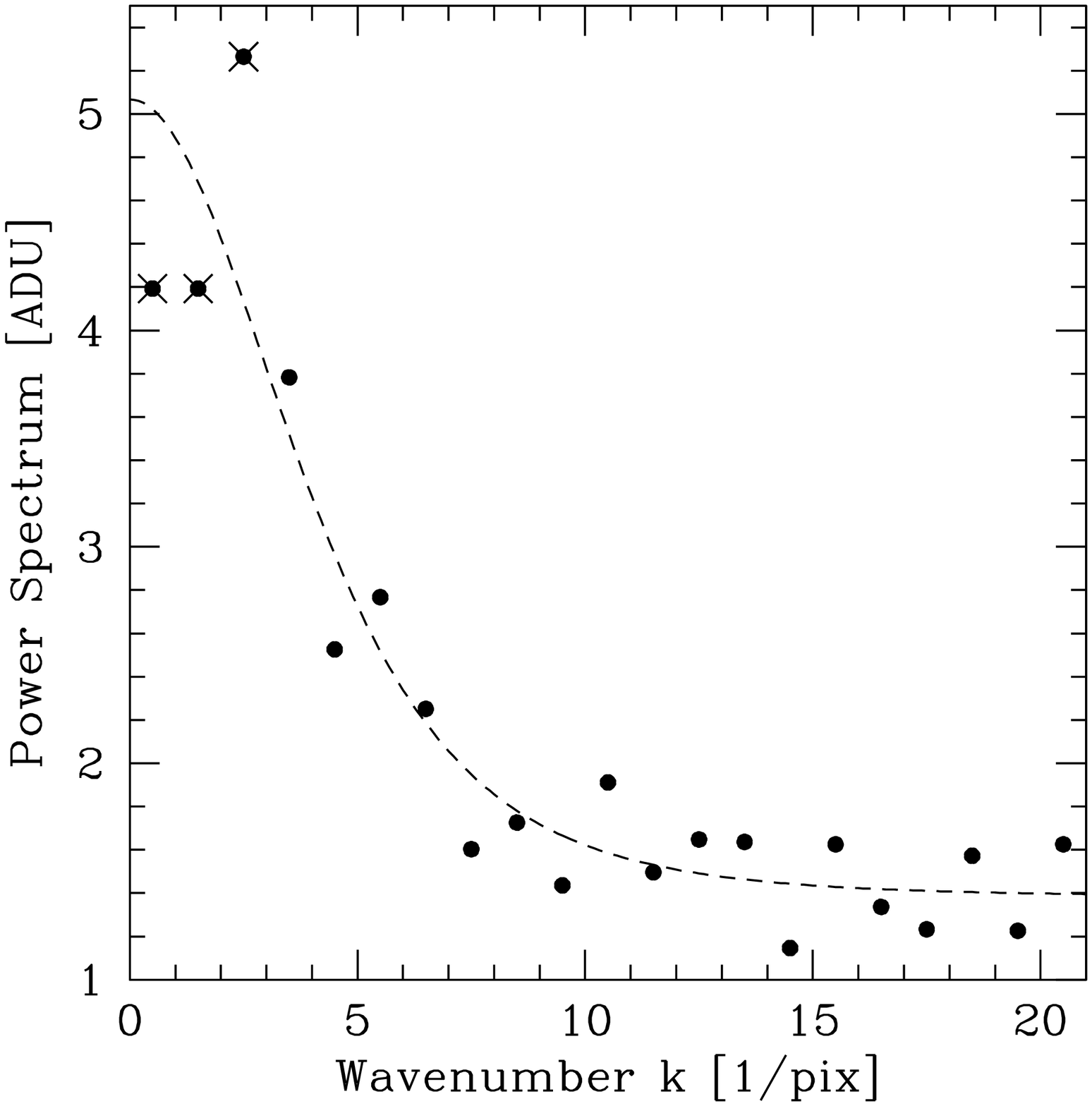,height=3.6cm,width=3.8cm}\hspace{0.2cm}
\epsfig{figure=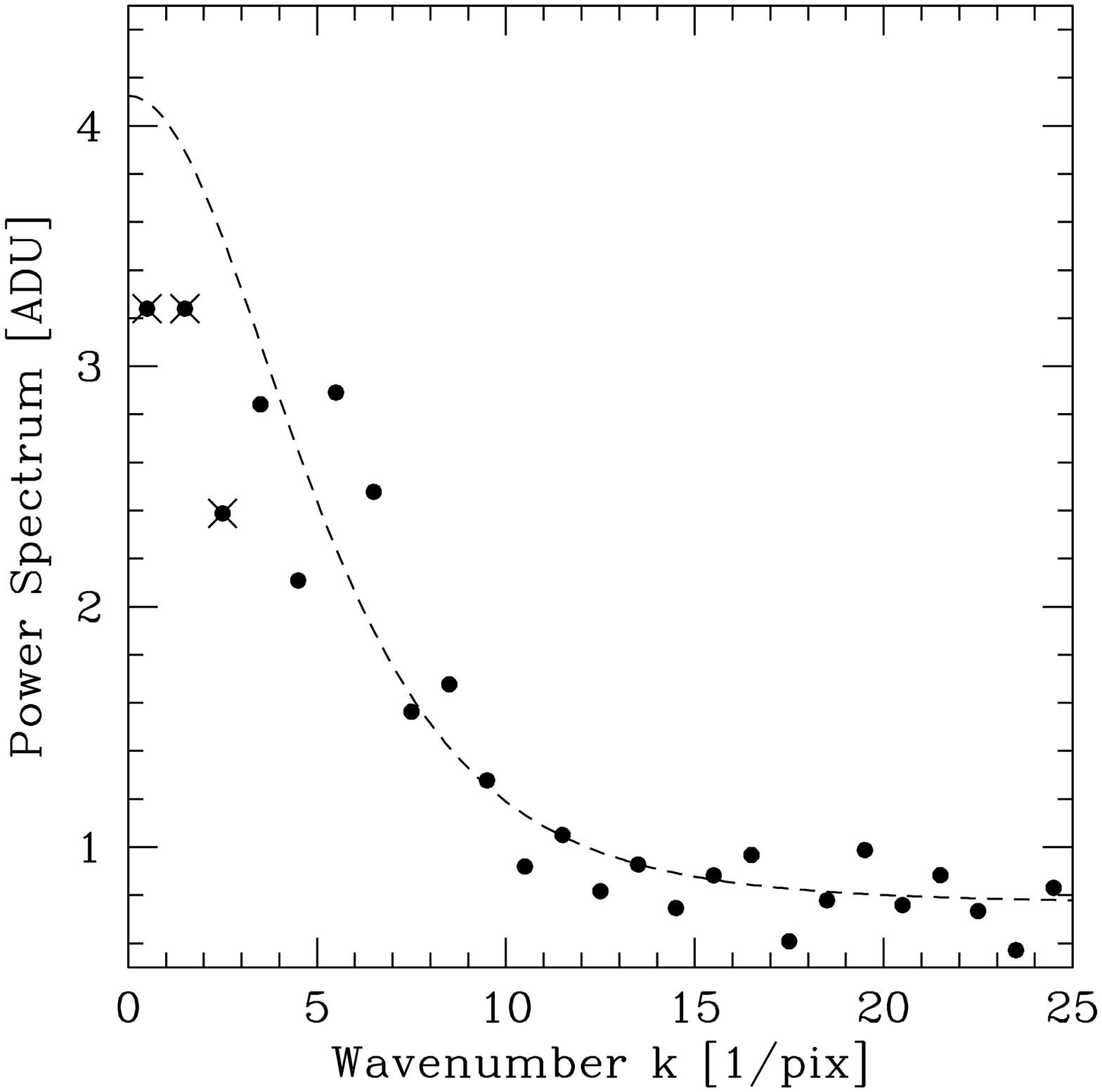,height=3.6cm,width=3.8cm}\hspace{0.2cm}
\epsfig{figure=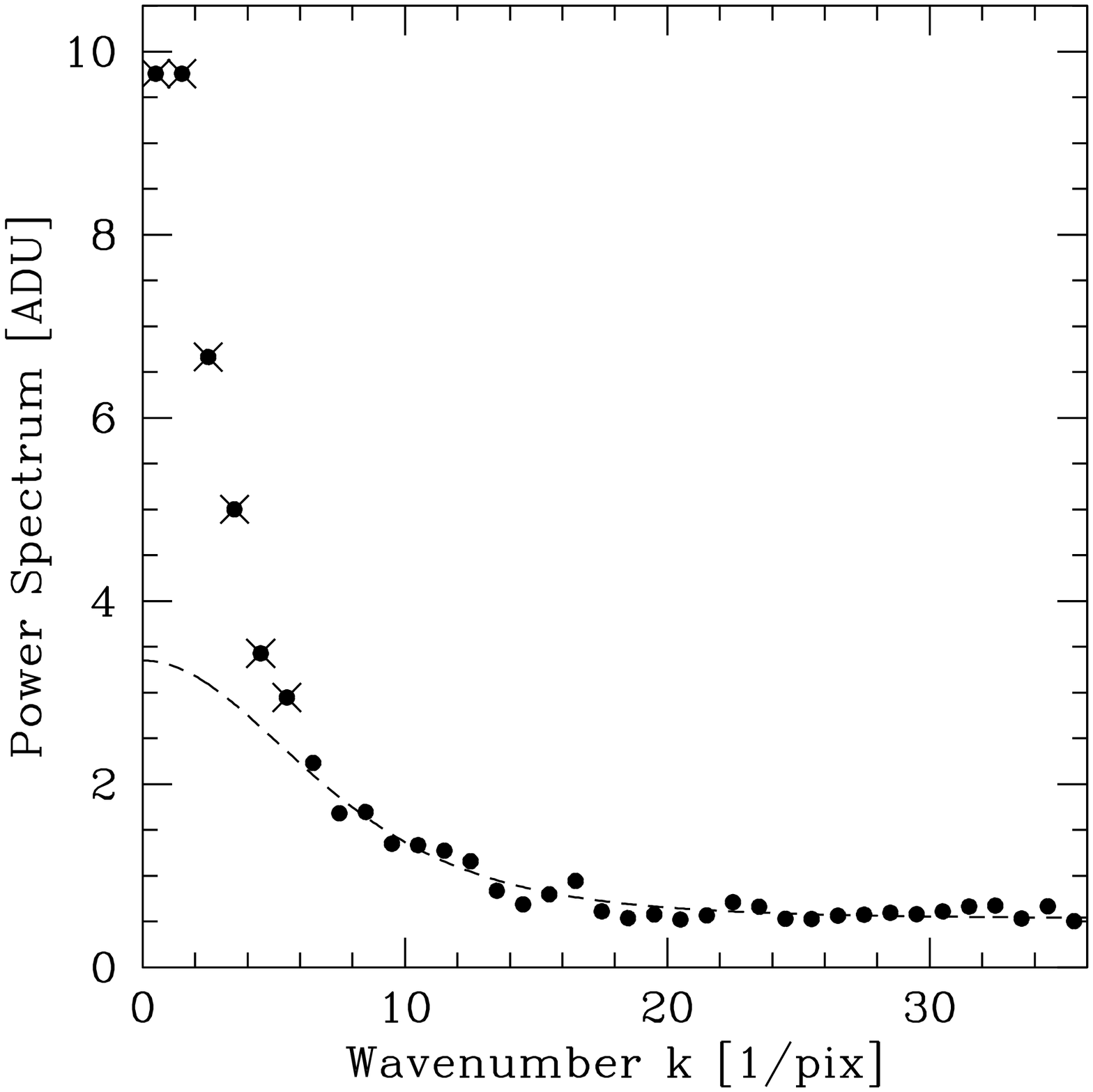,height=3.6cm,width=3.8cm}\hspace{0.2cm}
\epsfig{figure=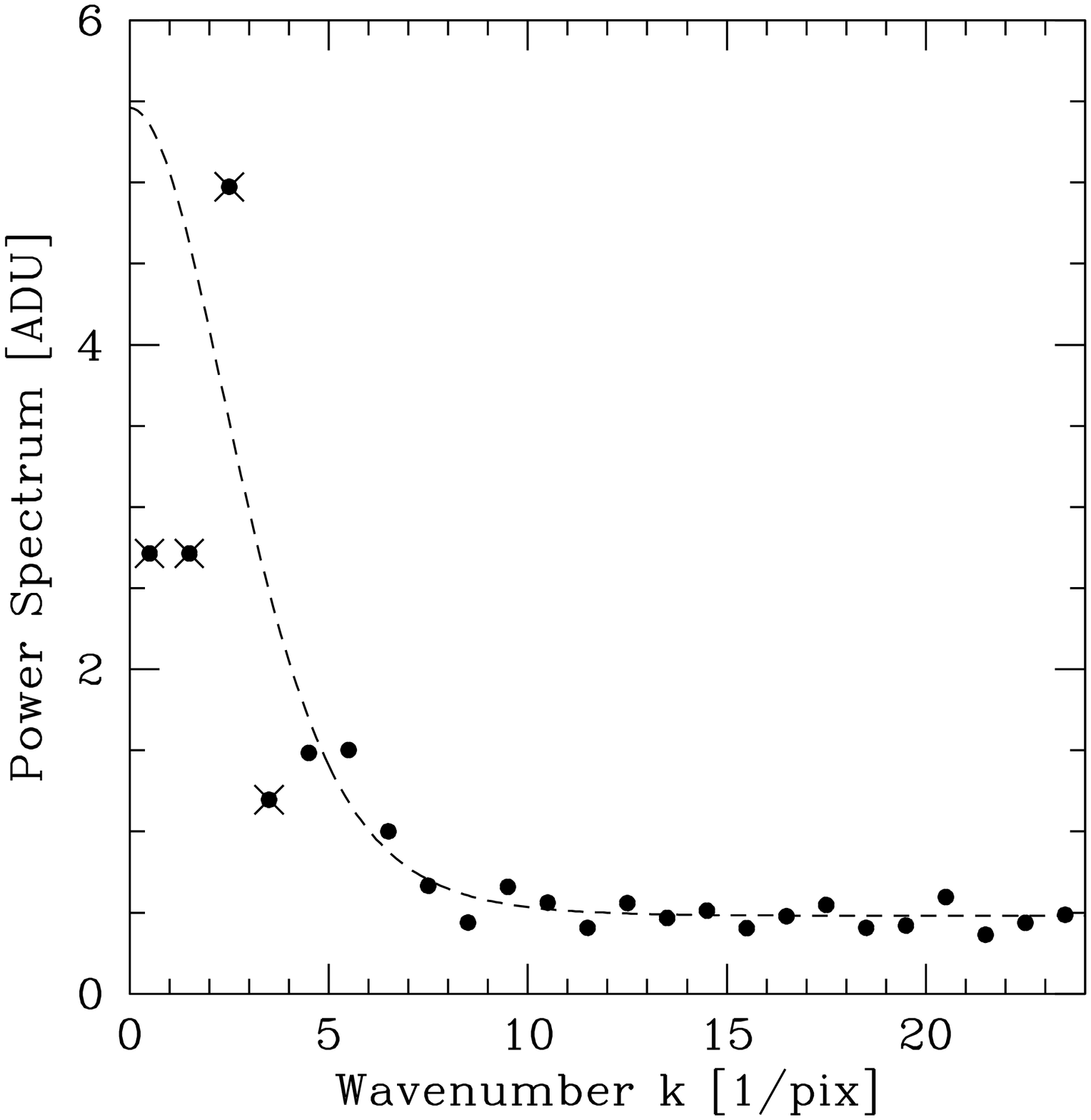,height=3.6cm,width=3.8cm}\\
\end{center}
\caption[]{\label{examples}Example images and plots for four galaxies, illustrating the 
SBF measurement procedure. From left to right: CCC 115 (field 3); CCC 75 (field 1); 
CCC 61 (field 1); CCC 125 (field 4). From top to bottom: Original galaxy image; elliptical
 model subtracted from former image; former image divided by square root of model, with 
contaminating sources and region outside measurable circle masked; power spectrum of the 
former image; azimuthal average of the former image with dashed line showing the result 
of the fit and crosses marking the points rejected for the measurement.}
\end{figure*}

The aim of the SBF measurement is the determination of the apparent fluctuation magnitude
 $\overline{m}_{\rm I}$. From $(V-I)_0$ one then derives $\overline{M}_{\rm I}$ via 
equations~(\ref{sbfrel}) and~(\ref{sbfrel3}) and thereby distance modulus. The following 
steps have been performed to measure $\overline{m}_{\rm I}$, see as well {\it MieskeI}:\\
1.~Model mean galaxy light with ELLIPSE using a sigma clipping algorithm to disregard
contaminating sources hidden below the galaxy light, subtract the model.\\
2.~Detect and subtract remaining contaminating objects from original image.\\
3.~Model mean galaxy light on the cleaned image.\\
4.~Subtract model of original image.\\
5.~Divide resulting image by square root of the model, cut out circular portion 
with radius typically 20 pixel ($4\arcsec$), corresponding to about 8 seeing disk diameters.\\
6.~Mask out contaminating sources like foreground stars, background galaxies and globular
clusters. The completeness limit of 
the contaminating source detections was determined by artificial star experiments
 using SExtractor and the 
ARTDATA package under IRAF. The limiting magnitude for point sources 
was about 25 mag in $I$.\\
7.~Calculate the power spectrum (PS) of the cleaned image.\\
8.~Obtain the azimuthal average of the PS.\\
9.~Fit function of the form 
\begin{equation}
\label{azimut}
P(k)=PSF(k)\times P_{\rm 0}+P_{\rm 1}
\end{equation}
to the azimuthally averaged PS.
$PSF(k)$ is the PS of the seeing profile, normalized to unity at k=0. It is determined
from a single star with no close neighbours in the same frame by fitting a Moffat profile
to its PS. $P_{\rm 1}$ is the white noise component.
$P_{\rm 0}$ is the amplitude of the pixel-to-pixel surface brightness fluctuations.
We define as the signal-to-noise of the measurement
 $S/N\equiv P_{\rm 0}/P_{\rm 1}$. Values at small $k$ (long wavelength) are rejected for 
the fit, as they are often influenced by large scale residuals from imperfect galaxy 
subtraction (for more details see 
{\it MieskeI}).\\
10.~It holds for the desired observable $\overline{m}_{\rm I}$:
\begin{equation}
\label{mbarI}
\overline{m}_{\rm I}=-2.5*log(P_{\rm 0}) + ZP - A_I - \Delta k + \Delta GC -\Delta sim
\end{equation}
with $ZP$ being the photometric zero point including exposure time. $A_I$ is the foreground absorption, 
$\Delta k=z\times 7$ the k-correction for SBF in the $I$-band (Jensen et al. \cite{Jensen98}), 
$\Delta GC$ the contribution to the fluctuations caused by Globular Clusters (GCs) 
below the detection limit and $\Delta sim=0.15 \pm 0.05$ mag the bias correction that needs to 
be applied following the results of our SBF-simulations from {\it MieskeI}. 
Before going further,
we treat in some more detail the two corrections $\Delta sim$ and $\Delta GC$. 
As already shown in {\it MieskeI}, the fluctuations 
caused by background galaxies are negligible at the given depth of our data.\\
In Fig.~\ref{examples} thumbnail images of 4 investigated galaxies are given, illustrating 
the reduction procedure until obtaining the azimuthally averaged power spectrum.\\
\subsubsection{Including the bias found in simulations}
\label{bias}
In {\it MieskeI} we have presented SBF-simulations for artificial dEs at distances between 29.4 and 
33.4 mag, including varying seeing and integration times. The simulations are based on VLT-FORS1 zero
points. Their main purpose was to establish realistic magnitude limits
down to which the SBF-method can be applied in order to establish the membership of a candidate
dE in a galaxy cluster, depending on seeing and integration times.\\
A useful byproduct of these simulations are the error estimates for measuring $P_{\rm 0}$,
and bias estimates between input and output $P_{\rm 0}$. In {\it MieskeI}
we find a systematic bias of $\Delta sim=0.15 \pm 0.05$ mag between input and output SBF-magnitude 
towards measuring too faint SBF, which is independent on the simulated galaxy distance. We show
that the implementation and measurement of {\it not} seeing convolved pixel-to-pixel SBF yields no bias
at all and that convolution with the seeing yields less than 1\% flux loss. From that it is deduced that
recovering pixel-to-pixel fluctuations from seeing convolved images is 
subject to small, but non negligible loss of fluctuation signal in our simulations. 
We use the same SBF-measurement
procedure both for our real data presented here and for the simulations in {\it MieskeI}. 
Therefore, we apply the bias estimates found in the simulations to our results.\\
\subsubsection{Contributions from Globular Clusters}
To calculate $\Delta GC$, equation~(10) from Jensen et al. (\cite{Jensen98}) is used. 
-8.5 mag is assumed as the absolute 
$I$ band turnover magnitude TOM (Kundu \& Whitmore \cite{Kundu01}) of the GC luminosity function.
Apart from the approximate distance of the investigated galaxy, 
the most important ingredient of Jensen's equation (10) is the specific frequency $S_N$ of GCs,
defined as the number of GCs per $M_V=-15$ mag galaxy luminosity.
For all investigated galaxies except for
the giants NGC 4696 and NGC 4709, $S_N$ could not be 
measured precisely due to the low number counts involved: first, the expected total number of GCs 
$N_{GC}$ is of the order of one up to a few tens, as $M_V>-18$ mag and therefore $N_{GC}<16\times S_N$; 
second, only half of them can be detected due to 
the completeness limit. Therefore, we use the results of Miller et al. (\cite{Miller98}) 
on $S_N$ for dwarf elliptical galaxies and adopt a value of $S_N=4 \pm 3$ for all investigated 
dwarf galaxies plus the fainter giant CCC 89. The error in $\Delta GC$ is given by the error
range of $\pm 3$ for $S_N$. 
As at a given $S_N$, the distance $(m-M)$ of the galaxy 
must be known to correctly calculate $\Delta GC$, $\Delta GC$ and $(m-M)$ were determined 
iteratively. $\Delta GC$ for 
the different galaxies ranged between 0.02 and 0.28 mag with a mean of 0.11 mag. In general,
$\Delta GC$ is larger for redder than for bluer galaxies, as the strength of the SBF decreases
with redder colour while the assumed GC contribution remains equal.\\
\subsubsection{Error calculation for $(m-M)$}
\label{dmerr}
\begin{figure}
\begin{center}
\epsfig{figure=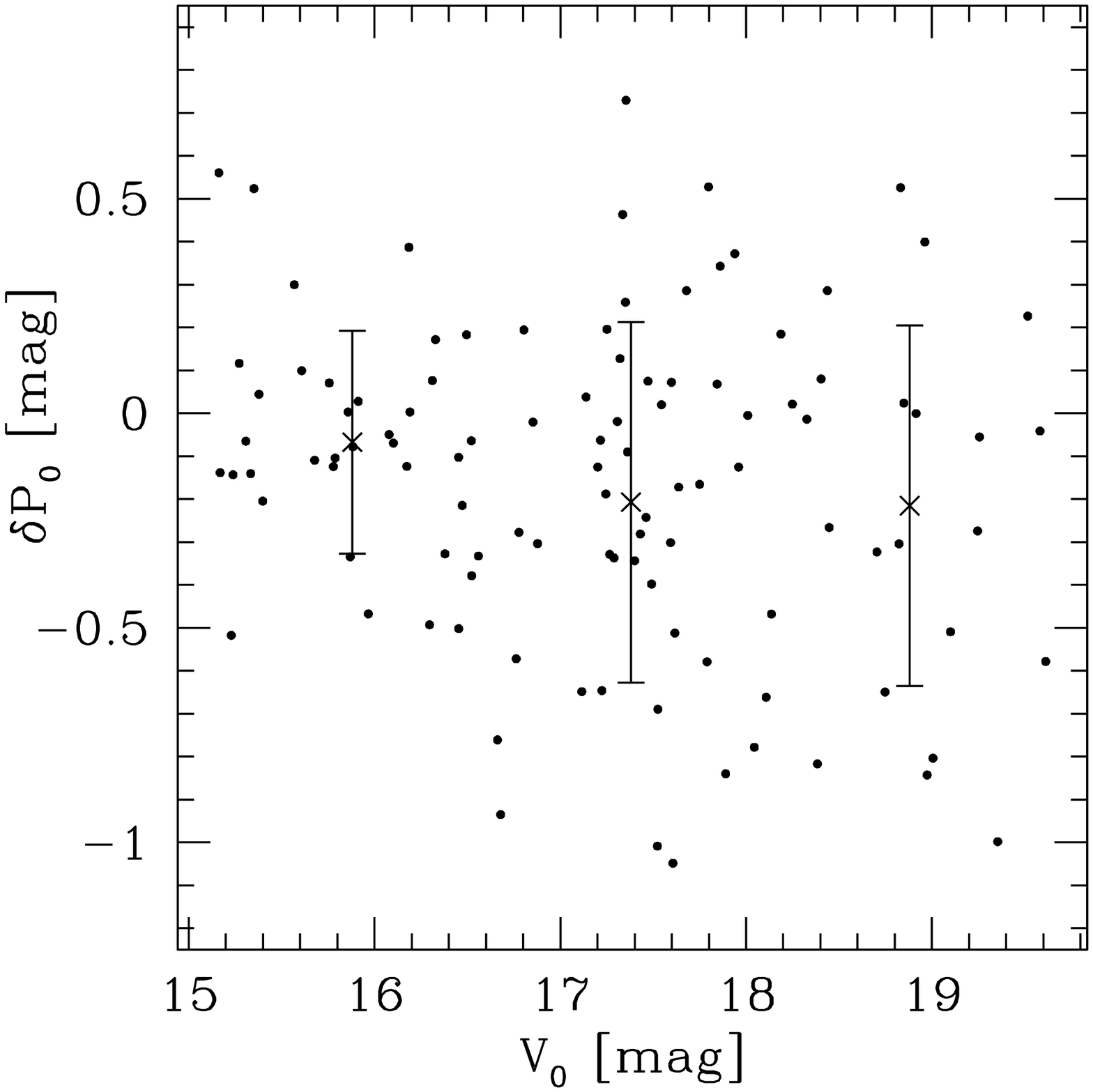,width=8.6cm}\\
\end{center}
\caption[]{\label{simcomp}Apparent magnitude $V_0$ of simulated dEs at 32.8 mag distance modulus
(taken from Fig.~8 of {\it Mieske I} for $V_0<17.7$ mag, new simulations otherwise) plotted 
vs. the measurement deviation $\delta P_0$ between
simulated and measured SBF-magnitude. The y-positions of the three error bars indicate the 
mean $\delta P_0$ in the
range $\pm$ 0.75 mag of their x-position. The size of the error bars corresponds to the rms-scatter
around the mean  $\delta P_0$ in that magnitude range.}
\end{figure}

The error of $(m-M)$ consists of the measurement error $\delta\overline{m}_I$ and the derivation
error $\delta \overline{M}_I$.\\
$\delta\overline{m}_I$ is composed by the single errors of 
the different terms from equation~(\ref{mbarI}). The most important error contribution
here comes from $P_{\rm 0}$. We derive the error of $P_{\rm 0}$ from Monte Carlo simulations, using 
the scatter of $P_{\rm 0}$ measurements on simulated dEs from {\it Mieske I}, complemented for fainter
magnitudes by new simulations. 
In Fig.~\ref{simcomp}, apparent magnitude $V_0$ is plotted vs. SBF measurement deviation $\delta P_0$
for 108 simulated dEs at the approximate Centaurus cluster distance. The
magnitude range $15.1<V_0<19.6$ was chosen to cover the same range occupied by the real galaxies
in this paper,
excluding the magnitude regime of the two giants NGC 4709 and NGC 4696 (see Sect.~\ref{gcs}).
We subdivide the simulated magnitude range in three bins of 1.5 mag width. 
Within each bin, both the mean $\delta P_0$ 
and the rms-scatter around it is indicated in Fig.~\ref{simcomp}. The scatter
is 0.26 mag for the brightest bin and 0.42 mag for both fainter bins. We adopt
these rms-scatters in the different bins as the error in $P_{\rm 0}$ for our measurements.
The corresponding mean S/N of the SBF-measurement ranges between 3 for the faintest bin and 9 for
the brightest one. We apparently do not deal with high S/N data nor with large sample
areas for SBF measurement, which explains the considerable
uncertainty in the fainter magnitude bins, corresponding to about 20\% in distance. The slightly
larger mean for the brightest bin is caused by the stronger contribution from undetected GCs
to the SBF-signal, as the brighter the galaxies the redder the colour and the weaker the SBF-signal.
Note that the means of $\delta P_0$ agree with the adopted 0.15 mag bias correction 
(see Sect.~\ref{bias}).\\
\noindent $\delta \overline{M}_I$ is composed by
the cosmic scatter of 0.10 mag and the error in measuring $(V-I)$ multiplied by the slope
of equations~(\ref{sbfrel}) or~(\ref{sbfrel3}), respectively.\\ 
\subsubsection{Special treatment for NGC 4696 and NGC 4709}
\label{gcs}
\begin{figure*}
\begin{center}
\epsfig{figure=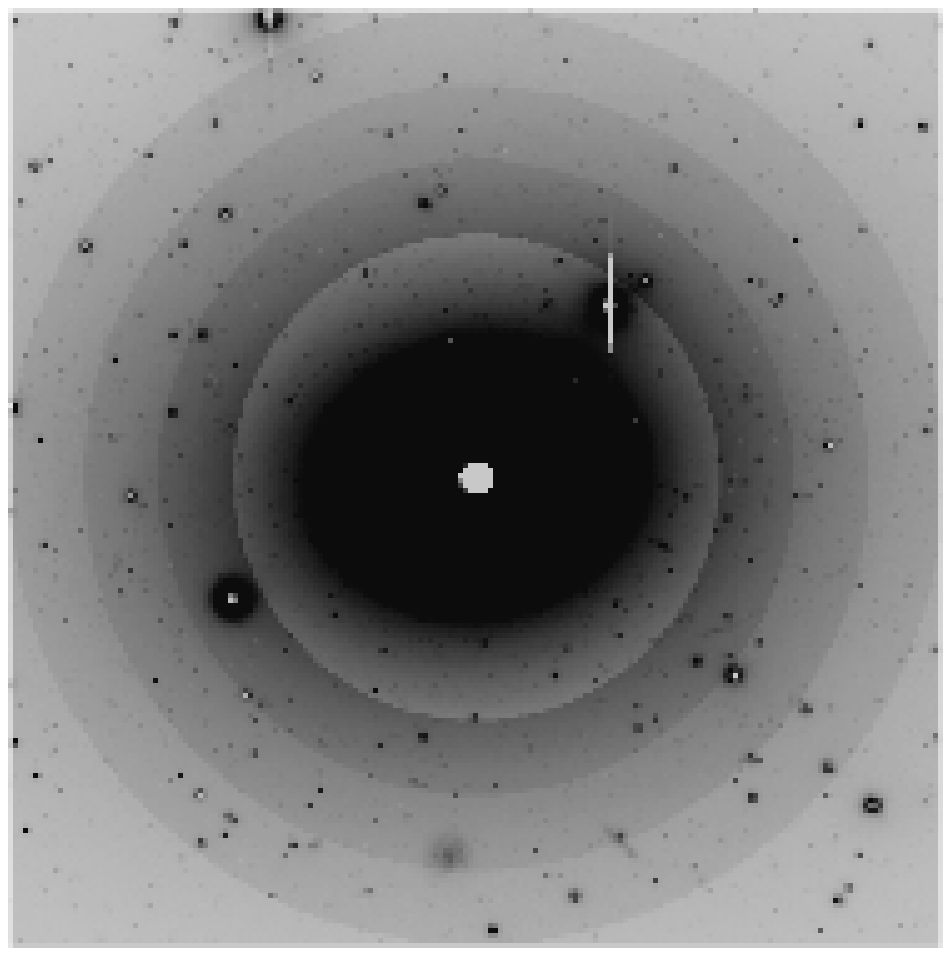,height=6.6cm,width=6.7cm}\hspace{0.2cm}
\epsfig{figure=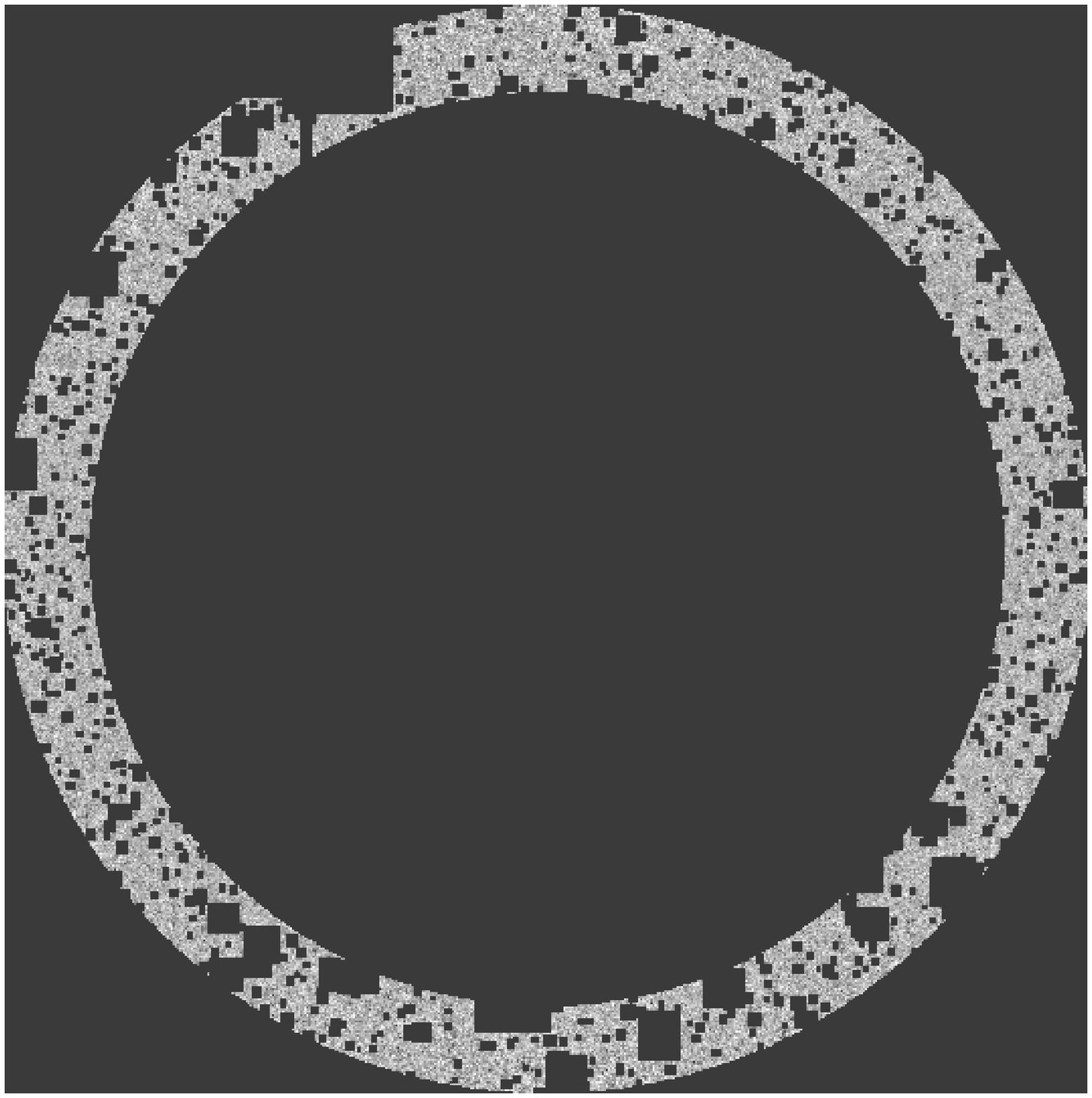,height=6.6cm,width=6.7cm}\\\hspace{-0.05cm}
\epsfig{figure=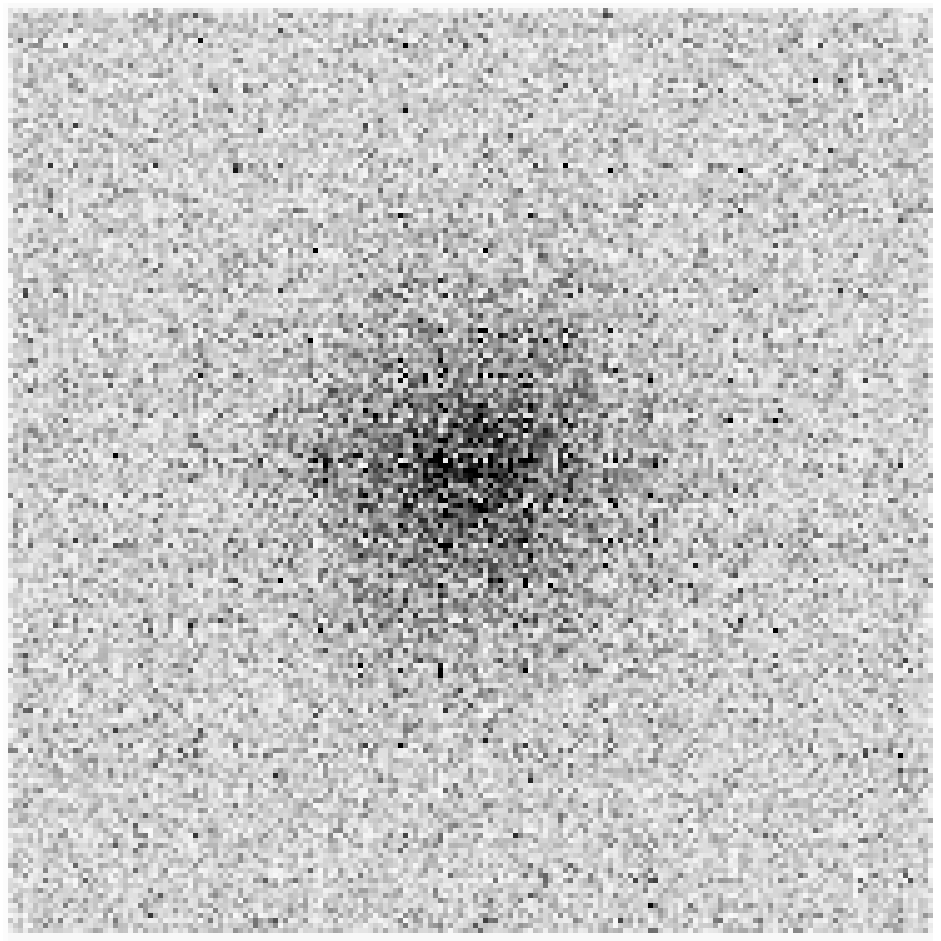,height=6.4cm,width=6.8cm}\hspace{0.2cm}
\epsfig{figure=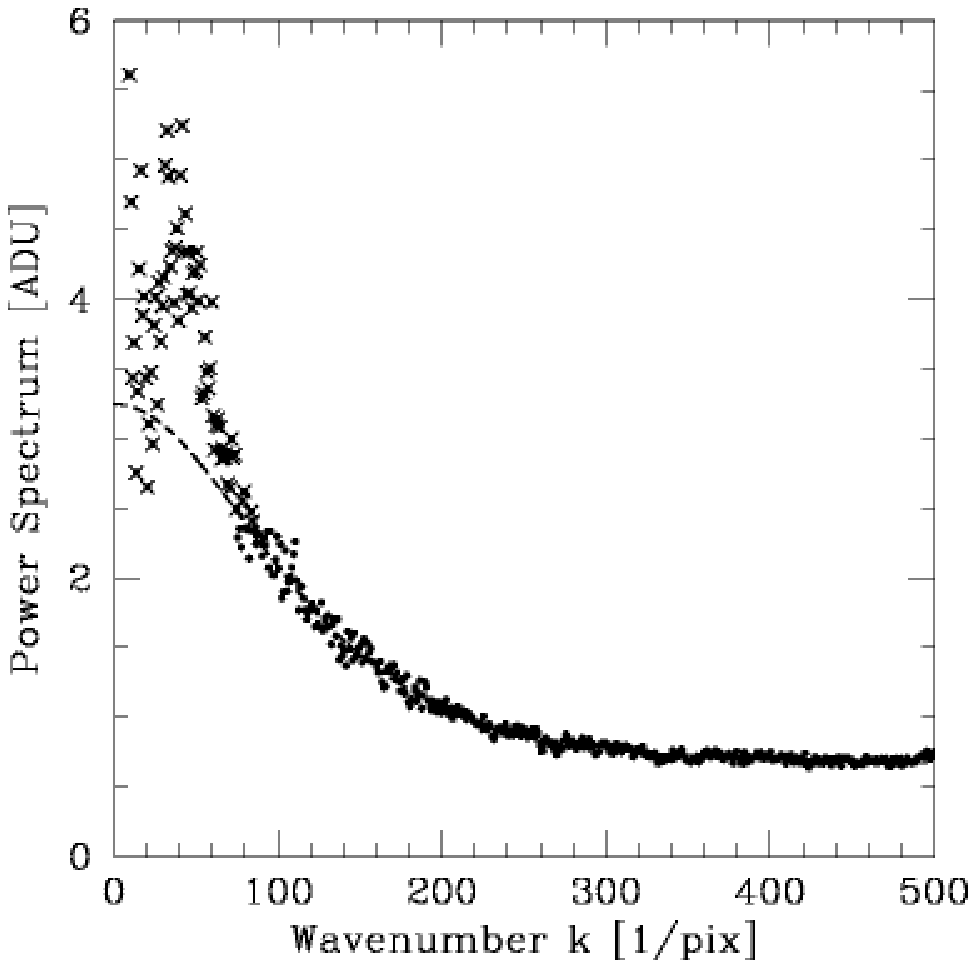,height=6.3cm,width=6.8cm}\\
\end{center}
\caption[]{\label{n4696}Example images and plots illustrating the SBF measurement 
procedure for NGC 4696, the central galaxy of the Centaurus cluster. From left to right 
and top to bottom: 1. Original galaxy image. The three rings within which SBF were 
measured are indicated by different offset intensities applied to the respective rings. 2. 
Image containing only the outermost ring, with any contaminating objects masked. 
3. two-dimensional power spectrum of image 2. 4. Azimuthally 
average of the former image with dashed line showing the result of the fit and crosses 
marking the points rejected for the measurement.}
\end{figure*}

The SBF measurement procedure was slightly different for the two bright and extended 
giants NGC 4696 and NGC 4709.\\
The first difference was that SBF were measured in adjacent {\it rings} centered 
on the galaxy, not in circles, leaving out the central part. This was necessary, 
because in their innermost part the Poisson noise caused by the high surface brightness ($\mu_I\simeq$ 
18 mag/arcsec$^2$) gives a considerable 
shift to the bright of the completeness magnitude for detecting GCs, resulting in a 
large uncertainty when calculating the contribution from undetected GCs. To avoid that, 
the inner galaxy parts with intensity at least half that of the sky were disregarded 
for the SBF measurement (see as well Tonry \& Schneider \cite{Tonry88}). Besides,
NGC 4696 has prominent dust lanes close to its center which do not allow SBF measurement. 
The final value of $P_0$ for NGC 4696 and NGC 4709 corresponds to the mean value 
obtained in the different rings after scaling $P_0$ according to the difference 
between $(V-I)$ in the respective ring and the mean $(V-I)$ of all rings. 
This is done 
as a consequence of the colour-SBF relation. The error is adopted as
the standard deviation of the different values. 
For NGC 4696, the SBF were measured in 3 rings ranging 
between 52 and 100$''$ from the galaxy center, for NGC 4709 in 2 rings between 32 and 54$''$. 
See Fig.~\ref{n4696} as an example. Unlike for the fainter rest of the investigated
galaxies, the error of $P_0$ was only of the order of a few percent 
(see as well Table~\ref{resultstab}). This is because the area used to measure SBF was several hundred 
times larger.\\
The second difference was that the specific frequency $S_N$ could 
be calculated more precisely than the rough estimate adopted for the smaller, mainly dwarf 
galaxies. To do so, we obtained aperture photometry of all sources in the rings where SBF 
were measured. To select GCs, we demanded the sources to be unresolved, fainter than $I$=21 
mag and to be in the colour range $0.8<(V-I)<1.3$ mag (Kissler-Patig et al. \cite{Kissle97}). 
By doing the same photometry and 
applying the same selection criteria to a background field we calculated the number density 
of contaminating background objects, which proved to be negligible. The simultaneous 
detection incompleteness for GCs in $V$ and $I$ reached 50\% at typically $I$=24.5 mag, a 
limit about 0.5 mag brighter than for detection only in the $I$-band. A Gaussian was fit to 
the incompleteness corrected GC luminosity function until the 50\% completeness limit. The 
error of the fitted parameters turn-over magnitude (TOM) and total number of GCs was 
calculated by varying the width $\sigma$ by $\pm$ 0.2 mag around a mean of 1.3 mag. This was 
done due to the well known fact that TOM and $\sigma$ are covariant. 1.3 mag is the mean 
$\sigma$ from Kundu \& Whitmore's (\cite{Kundu01}) HST results on the GCSs of a large 
sample of early-type galaxies, and 0.2 mag is the standard deviation of their results.\\ 
The results of our GCLF measurements are given in table \ref{resultsgcs}. For both galaxies,
the TOMs are 
brighter than the incompleteness limit. For NGC 4696, the derived 
$S_N$ is 4.4 $\pm$ 0.8, for NGC 4709 it is 2.1 $\pm$ 0.5. This results in 
$\Delta GC$=0.08 for NGC 4696 and $\Delta GC$=0.02 for NGC 4709 (see Table~\ref{resultstab}). 
Note that the values for $S_N$ refer only to the ring-regions where SBF were measured, not
to the entire galaxies. For NGC 4696, our value is consistent with the result of
Lee \& Geisler (\cite{Geisle97} and Lee, private communication), who obtain $S_N=$ 6 $\pm$ 1.
For NGC 4709, our result is consistent with the mean $S_N=2.4 \pm 1.8$ 
obtained by Kundu \& Whitmore (\cite{Kundu01})
from HST investigations of 28 nearby early type giants.\\
In a forthcoming paper (Hilker \& Mieske, in prep.), 
the GC systems of NGC 4696 and NGC 4709 will be dealt with in more detail.\\
\begin{table*}
\begin{center}
\begin{tabular}{r|lllllll}
Gal-Nr. & TOM [mag]& $(m-M)_{GC}$ & $(m-M)_{SBF}^*$ &$I_{\rm cut}$ [mag] &N$_{\rm GC, rings}$& $M_{\rm V, rings}$ & $S_{\rm N, rings}$\\\hline
 NGC 4696 & 24.25 $\pm$ 0.2 & 32.75 $\pm$ 0.2 & 32.84 $\pm$ 0.14 & 24.5 & 1140 $\pm$ 110 & $-$21.05 $\pm$ 0.2 & 4.4 $\pm$ 0.8\\
 NGC 4709 & 23.6 $\pm$ 0.2 & 32.1 $\pm$ 0.2 & 32.36 $\pm$ 0.15 & 24.6 & 143 $\pm$ 20 & $-$19.60 $\pm$ 0.2 & 2.1 $\pm$ 0.5\\
\end{tabular}
\end{center}
\caption[]{\label{resultsgcs}Details of the $S_N$ measurement for the two giant Centaurus 
members NGC 4696 and NGC 4709. Note that this measurement is restricted only to the
rings where SBF were measured. $(m-M)_{GC}$ is calculated assuming -8.5 mag as the absolute 
$I$ band TOM (Kundu \& Whitmore \cite{Kundu01}).
$I_{\rm cut}$ is the limiting magnitude for the GCLF fitting. 
N$_{\rm GC, rings}$ is the incompleteness corrected total number of GCs in the rings where SBF were 
measured. $M_{\rm V, rings}$ is the absolute magnitude in the same region. For its calculation the 
distance modulus adopted was the mean of the SBF and GCLF distance modulus. \hspace{1.2cm} 
$^*$From Table~\ref{resultstab}.}
\end{table*}
\section{Results}
\begin{figure}
\begin{center}
\epsfig{figure=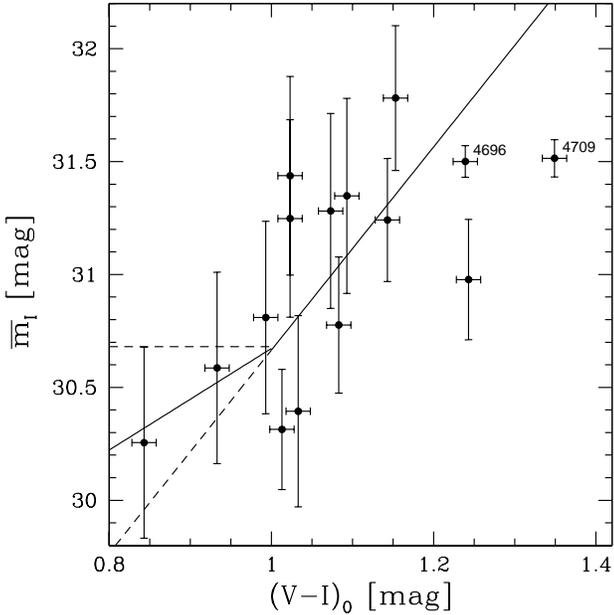,width=8.6cm}\\
\end{center}
\caption[]{\label{visbf}Colour-SBF diagram of the investigated Centaurus cluster galaxies.
The location of NGC 4696 and NGC 4709 is indicated by their
NGC number. 
The solid line corresponds to equations~(\ref{sbfrel}) ($(V-I)>1.0$) and~(\ref{sbfrel3}) 
($(V-I)\le 1.0$) 
for a distance modulus of (m-M)=33.08 mag, the mean distance of the investigated galaxies. 
Dashed lines as in Fig.~\ref{visbftheo}. The errors in $(V-I)$ are estimated from the
uncertainty in the local sky level determination for $V$ and $I$. For NGC 4696 and NGC 4709,
the errors in $\overline{m}_{\rm I}$ are
 estimated from the scatter in $\overline{m}_{\rm I}$ between the different investigated rings
(Sect.~\ref{gcs}). For the other galaxies they are estimated from Monte Carlo simulations 
(Sect.~\ref{dmerr}).}
\end{figure}

\begin{figure}
\begin{center}
\epsfig{figure=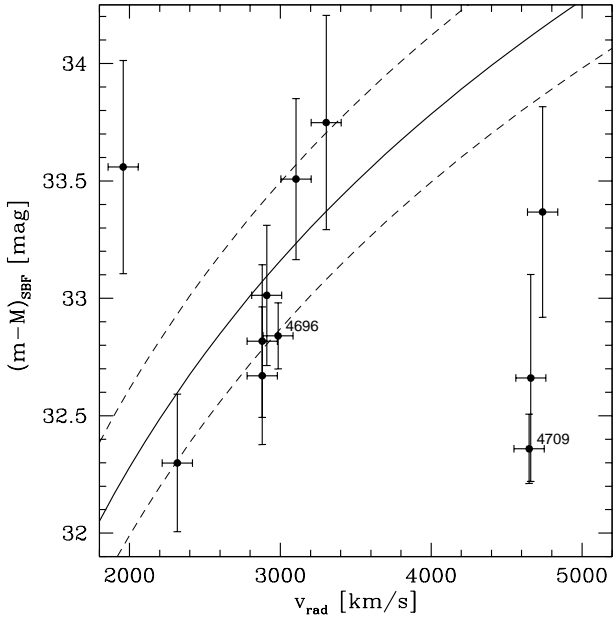,width=8.6cm}\\
\end{center}
\caption[]{\label{raddm}Heliocentric radial velocity is plotted vs. distance modulus for 
the investigated galaxies with measured redshifts. The location of 
NGC 4696 and NGC 4709 is indicated by their NGC number. 
The solid line gives the Hubble flow 
for $H_{\rm 0}$=70 km/s/Mpc. The upper dashed line corresponds to $H_{\rm 0}$=60, the 
lower dashed line to $H_{\rm 0}$=80. The fact that the data points are not aligned 
along these lines but rather show no correlation indicates
that the galaxies are either bound within one cluster or are situated in a region with gravitational 
distortion.}
\end{figure}

\begin{figure}
\begin{center}
\epsfig{figure=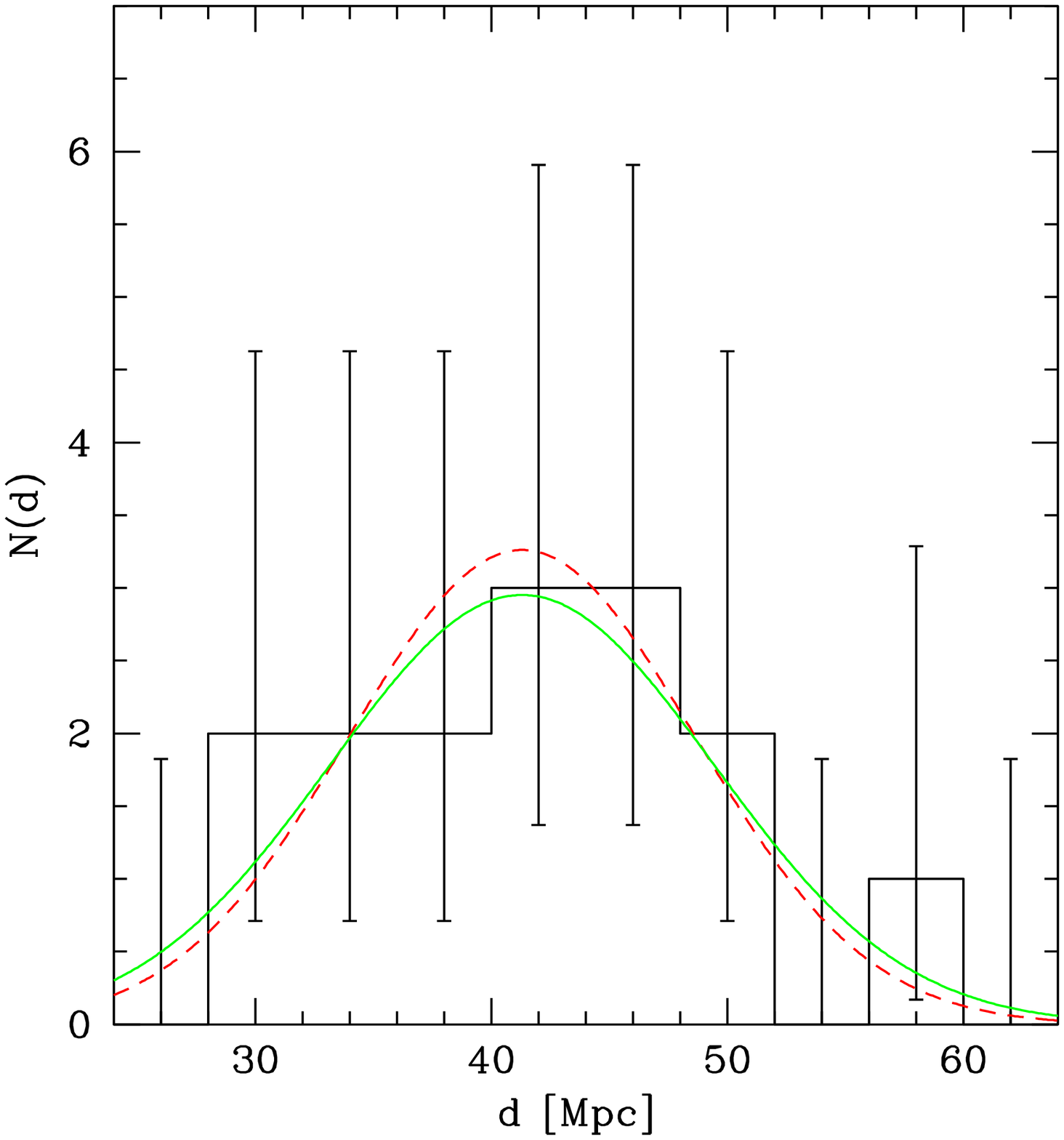,width=8.6cm}\\
\end{center}
\caption[]{\label{dmhist}Histogram of the SBF distances for the investigated Centaurus 
Cluster galaxies. Note that the error bars do not correspond to the square root of the bin 
value, but are calculated based on the formulae given by Gehrels (\cite{Gehrel86}) 
for the calculation 
of errors for low number statistics.
The solid line represents a gaussian distance 
histogram centered on the mean 41.3 Mpc with a half-width of 8.1 Mpc, 
corresponding to the standard deviation of the measured distances around their mean. 
The dashed line represents a gaussian distance histogram with half-width 7.3 Mpc, 
which is the mean single distance measurement uncertainty (see text and 
Table~\ref{resultstab}).}
\end{figure}

The results of all SBF measurements are summarized in Table~\ref{resultstab} and plotted in 
Figures~\ref{visbf}~to~\ref{dmhist}. To calculate the k-correction for the galaxies with 
no measured radial velocity available in the literature, 3000 km/s was assumed.
The error of $(m-M)$ in the table is the quadratic sum of the error in measuring 
$\overline{m}_{\rm I}$ and in deriving $\overline{M}_{\rm I}$ from $(V-I)$ 
(see Sect.~\ref{dmerr}). The error in metric distance $d$ is the mean of the 
upper and lower distance error range
corresponding to the magnitude error in $(m-M)$.\\
\begin{table*}
\begin{center}
\begin{tabular}{rrrrrrrrrr}
CCC-Nr. & Field & $P_0$ [ADU] & $P_1$ [ADU]& $ZP$ & $S/N$ & $\Delta GC$ & $\overline{m}_{\rm I}$ & $(m-M)$ & $d$ [Mpc]\\\hline
52 &  1 & 2.81 $\pm$ 1.18 & 0.64 & 32.76 & 4.39 & 0.15 & 31.35 $\pm$ 0.43 & 33.34 $\pm$ 0.45 & 46.7 $\pm$ 9.7\\
61 &  1 & 2.94 $\pm$ 0.76 & 0.51 & 32.76 & 5.77 & 0.09 & 31.24 $\pm$ 0.27 & 33.01 $\pm$ 0.30 & 40.0 $\pm$ 5.5\\
65 (N4696) &  1 & 2.33 $\pm$ 0.04 & 0.49 & 32.78 & 4.78 & 0.08 & 31.50 $\pm$ 0.07 & 32.84 $\pm$ 0.14 & 37.0 $\pm$ 2.4\\
70 &  1 & 3.56 $\pm$ 0.93 & 0.32 & 32.76 & 11.12 & 0.02 & 30.98 $\pm$ 0.27 & 32.30 $\pm$ 0.29 & 28.8 $\pm$ 3.9\\
75 &  1 & 3.30 $\pm$ 1.39 & 0.86 & 32.77 & 3.84 & 0.19 & 31.48 $\pm$ 0.44 & 33.56 $\pm$ 0.45 & 51.5 $\pm$ 10.9\\
89 &  2 & 2.16 $\pm$ 0.56 & 0.29 & 32.78 & 7.45 & 0.28 & 31.78 $\pm$ 0.32 & 33.51 $\pm$ 0.34 & 50.3 $\pm$ 8.0\\
111 &  3 & 6.02 $\pm$ 1.56 & 0.46 & 32.67 & 13.09 & 0.03 & 30.31 $\pm$ 0.27& 32.67 $\pm$ 0.29 & 34.2 $\pm$ 4.6\\
115 &  3 & 4.02 $\pm$ 1.69 & 1.34 & 32.67 &  3.00 & 0.09 & 30.81 $\pm$ 0.43& 33.24 $\pm$ 0.44 & 44.5 $\pm$ 9.2 \\
121 &  3 & 2.65 $\pm$ 1.11 & 0.76 & 32.67 & 3.49 & 0.15 & 31.28 $\pm$ 0.43& 33.37 $\pm$ 0.45 & 47.2 $\pm$ 9.8\\
123 &  3 & 5.38 $\pm$ 2.26 & 0.62 & 32.67 & 8.68 & 0.03 & 30.39 $\pm$ 0.42& 32.66 $\pm$ 0.44 & 34.1 $\pm$ 7.0\\
124 &  3 & 6.28 $\pm$ 2.64 & 3.36 & 32.67 & 1.87 & 0.02 & 30.26 $\pm$ 0.42& 33.02 $\pm$ 0.56 & 40.2 $\pm$ 10.6\\
130 (N4709) &  3 & 1.90 $\pm$ 0.10 & 0.38 & 32.67 &  5.00 & 0.02 & 31.51 $\pm$ 0.08 & 32.36 $\pm$ 0.15 & 29.6 $\pm$ 2.0 \\
125 &  4 & 5.28 $\pm$ 1.37 & 0.47 & 32.78 & 11.23 & 0.24 & 30.78 $\pm$ 0.30 & 32.82 $\pm$ 0.33 & 36.6 $\pm$ 5.5\\
58 &  5 & 2.62 $\pm$ 1.1 & 0.92 & 32.72 & 2.85 & 0.21 & 31.44 $\pm$ 0.44 & 33.75 $\pm$ 0.46 & 56.2 $\pm$ 11.9\\
68 &  6 & 5.03 $\pm$ 2.11 & 4.13 & 32.72 & 1.22 & 0.06 & 30.59 $\pm$ 0.42 & 33.15 $\pm$ 0.47 & 32.7 $\pm$ 9.2\\\hline\vspace{-0.1cm}\\
& & & & & & & &                   & 41.3 $\pm$ 2.1\\
\end{tabular}
\end{center}
\caption[]{\label{resultstab}Result of the SBF measurements for the investigated Centaurus 
cluster galaxies. $ZP$, $\Delta GC$, $\overline{m}_{\rm I}$ and $(m-M)$ are given in magnitudes.
The error in $(m-M)$ 
is the quadratic sum of the error in $\overline{m}_{\rm I}$ and in deriving $\overline{M}_{\rm I}$ 
from $(V-I)$ (see text). The distance error is the mean of the upper and lower distance error range
corresponding to the magnitude error in $(m-M)$. In the lowest row, the mean distance $d$ is given. }
\end{table*}
\label{results}
\subsection{Distance to the Centaurus cluster and its subcomponents}
The mean distance of all investigated galaxies is 41.3 $\pm$ 2.1 Mpc, corresponding to
33.08 $\pm$ 0.11 mag in distance modulus. Our result is higher than Tonry's result
derived from their SBF survey, which was 32.63 $\pm$ 0.09 mag based on SBF measurements of 
5 Cen30 and 3 Cen45 early-type giants ({\it SBF IV}). Although Dressler (\cite{Dressl93})
was the first to publish SBF-distances to Centaurus cluster galaxies, we will in the 
following compare
our results only with Tonry's newer values, as they have refined and complemented
Dressler's early measurements.\\
For the 2 galaxies common to 
both our and Tonry's data set, 
namely NGC 4696 and NGC 4709, Tonry et al. derive $(m-M)=32.75 \pm 0.17$ mag for NGC 4696 and 
$(m-M)=32.74 \pm 0.23$ for NGC 4709. To find out whether our results are consistent with that, 
we quadratically subtract the cosmic scatter error contribution of 0.1 mag from our distance error, 
as we compare the same galaxies. We then get
32.84 $\pm$ 0.10 mag for NGC 4696 and 32.36 $\pm$ 0.11 mag for NGC 4709. Tonry's and our distance for
NGC 4696 agree very well, while the distances for NGC 4709 differ by 1.1 sigma, or almost 0.4 mag.
A further discussion of the mean difference in distance between our and Tonry's 
Cen30 sample is given in Sect.~\ref{discussion}.\\
\subsubsection{Cen30 and Cen45 distance}
\label{Cen3045}
We separate our sample galaxies according to their radial velocity into Cen30 
($v_{rad}<4580$ km/s) and Cen45 ($v_{rad}>4580$ km/s), following Stein et al. 
(\cite{Stein97}).
We get 8 galaxies in Cen30 and 3 in Cen45. 
4 of our 15 galaxies do not have a radial velocity measured. In Fig.~\ref{raddm} it can be 
verified that there is no obvious distance separation between the two subsamples. The 
distance modulus corresponding to the mean distance of our 8 Cen30 galaxies is 33.11 $\pm$ 0.17 mag, 
of our 3 Cen45 galaxies it is 32.84 $\pm$ 0.29 mag. The distance moduli difference 
$(m-M)_{\rm Cen45}$ $-$ $(m-M)_{\rm Cen30}$ is
$-$0.27 $\pm$ 0.34 mag, consistent with both 
subclusters being at the same distance, but allowing for a considerable range of 
separations between 0.07 mag and $-$0.61 mag. This rules out that the subclusters 
 are separated by their Hubble flow distance of about $+$0.9 mag, and even favours
Cen45 being closer than Cen30.
Comparing our results with 
Tonry's 32.53 $\pm$ 0.11 mag for Cen30 and 32.81 $\pm$ 0.09 mag for Cen45 shows that only for
Cen45 are the results consistent within the error ranges. For Cen30, Tonry get a 0.58 mag shorter 
distance at a 2.1 $\sigma$ significance. We argue that this discrepancy is due to 
selection effects within the Tonry sample, as will be pointed out in more detail in 
Sect.~\ref{discussion}.\\
It is worth taking a special look at the distances to the two central galaxies of Cen30 (NGC 4696) 
and Cen45 (NGC 4709). Our measurements place NGC 4696 0.48 $\pm$ 0.21 mag 
more distant than NGC 4709, at a 2.3 $\sigma$ significance the distances are different.
The NGC 4696 distance is consistent with the mean distance of the whole sample and the mean 
distance of the 8 Cen30 galaxies. The NGC 4709 distance is only consistent with the Cen45 distance, 
but is shorter than the Cen30 distance and the distance of the whole sample. This might indicate that 
the distance difference found between the Cen30 and Cen45 sample is a real one. Due to the large
uncertainties involved in the distance measurement for the rest of our sample galaxies, the distance 
difference between the Cen30 and Cen45 central galaxies NGC 4696 and NGC 4709 
is a more precise indicator of the Cen30-Cen45 distance than the mean difference between the 
entire Cen30 and Cen45 sample, assuming that NGC 4696 and NGC 4709 are located
at the respective center of the two components.
The significant distance difference is supported by the investigations of 
their globular cluster systems (Sect.~\ref{gcs}), which show a difference of 0.65 $\pm$ 0.28 
mag between the two TOMs, placing NGC 4696 further than NGC 4709.\\
One can then interpret this separation
within the cluster-subcluster scenario such that Cen45 is a subgroup falling into Cen30 but 
not having reached it yet. In order not to base such a conclusion only on measurements
of the two main galaxies, SBF measurements from more giant Cen30 and Cen45 members are needed.\\
\subsection{$H_{\rm 0}$}
The well known peculiar velocity of the Local Group towards the GA of 300 $\pm$ 100 km/s
({\it SBF II}) allows a derivation of the
Hubble constant $H_0$ from our distance 
measurements.
Due to the large velocity dispersion of almost 1000 km/s observed for Cen30 galaxies 
(Stein et al. \cite{Stein97}), we prefer to adopt the mean heliocentric radial velocity 3170 $\pm$ 174
km/s of the 74 early type Cen30 galaxies investigated by Stein et al. rather than the mean 
2790 km/s of the 8 Cen30 members investigated by us, as the former velocity is 
much better defined because of its large underlying sample size.
The difference in mean radial velocity  between our sample and 
the Stein et al. sample is about 380 km/s. This difference lies well within the range of statistical
fluctuations, since with a sample of 8 galaxies
and a velocity dispersion of 1000 km/s, the accuracy of the mean is of the order of 350 km/s.
If we 
correct for the peculiar motion towards the GA, which we adopt to be 
at the location of the Centaurus cluster, 
we get 3470 $\pm$ 200 km/s as the mean Hubble flow velocity. The mean distance of the 
8 Cen30 members is 41.8 $\pm$ 3.4 Mpc. The resulting value for
the Hubble constant is then $H_0=$ 83.0 $\pm$ 8.3 km/s/Mpc.\\
The most precise current value for $H_0$ comes from the WMAP (Wilkinson Microwave Anisotropy 
Probe)-team (Bennett et al. \cite{Bennet03}, Spergel et al. \cite{Sperge03}), who 
give $H_0=71^{+4}_{-3}$ km/s/Mpc. Our value of 83.0 $\pm$ 8.3 km/s/Mpc agrees marginally
with theirs.\\
Assuming the WMAP value $H_0=$71 km/s/Mpc and the Cen30 distance of 41.8 $\pm$ 3.4 Mpc
derived by us, the undisturbed Hubble flow velocity at that distance would be 2970 $\pm$ 280 km/s. This
is remarkably consistent both with the mean heliocentric velocity of our 8 galaxies as well as of the
much larger sample of Stein et al. (\cite{Stein97}). It indicates that the peculiar velocities
of the Centaurus cluster galaxies with respect to the Hubble flow might be much smaller than 
previously found
by Tonry et al. ({\it SBF II}), which would result in smaller infall velocities into and hence a smaller
mass for the Great Attractor. This will be discussed in more detail in Sect.~\ref{cenga}.\\
\section{Discussion}
\label{discussion}
\subsection{Comparison with Tonry, Tonry's selection effects}
\label{seleff}
The distances to the two galaxies in common to both Tonry's and our data 
set agree to within their errors for NGC 4696 and differ with 
1.1 $\sigma$ significance for NGC 4709.
Our results place NGC 4709 0.48 $\pm$ 0.21 mag closer than NGC 
4696, while Tonry et al. obtain practically the same distance for both galaxies. The significant 
separation derived from our data is supported by the investigation of their GCSs.\\
We believe that the 
difference between our and Tonry's distance for NGC 4709 is caused by 
the different cutoff magnitudes for investigating the globular cluster systems. While we are 
able to map both GCSs well beyond their TOM (down to 25 mag in $I$), Tonry et al. have a 
significantly 
brighter dereddened cutoff-magnitude for their investigations, which is 23.6 $\pm$ 0.2 
mag for NGC 4696 and 
23.8 $\pm$ 0.3 mag for NGC 4709 in $I$ (Blakeslee \& Tonry, private communication). They obtain 
a contribution close to 50\% to the SBF-signal from undetected 
globular clusters, which causes an additional distance error of almost the same order 
(Blakeslee \& Tonry, private communication). The fact that NGC 4709 has a very poor GCS according to our deep 
data, could have made Tonry et al. overestimate the GC contributions for NGC 4709, as they 
only map it down to the TOM. In that case, they would have subtracted too much GC contribution from
their SBF signal, yielding a too weak SBF amplitude and consequently a too large distance.\\
An overall bias in the sense that Tonry's distances for their faintest survey galaxies might be 
too close has been 
discussed by {\it SBF IV} and Blakeslee et al. (\cite{Blakes02}), already. They argue that a 
combination of two factors makes them obtain too small distances at their faint survey limit, 
which was at about the Centaurus cluster redshift.\\
First, for a selection effect such that at 
a given cluster in which the SBF signal 
of the member galaxies is just at the limit of being detectable, one will only measure those 
whose observational errors place them above the detection limit and one will not measure those 
below the limit.
Second, a ``Malmquist bias'' 
(Malmquist \cite{Malmqu20}, Lynden-Bell \cite{Lynden88}), which referes to the distance bias 
arising from the spatial 
distribution of the sample galaxies, including the increase in the volume element with distance. 
This bias is porportional to the measured distance error and therefore stronger at the faint 
survey limits. Blakeslee et al. (\cite{Blakes02}) find that both of these effects are 
interrelated. They correct their measured distances for these selection effects
and obtain an overall correction of about 0.3 mag towards larger 
distances for the Centaurus cluster (Blakeslee \& Tonry, private communication). Thus, their 
mean Centaurus distance becomes about 32.9 mag, which agrees well with our result. Note
that the galaxy distances based upon which the mass of the Great Attractor and the corresponding
Hubble flow distortion 
was calculated in {\it SBF II} were not corrected for this bias, indicating a possible 
overestimation of peculiar velocities into the GA and hence its mass. See Sect.~\ref{cenga} for
further discussion.\\
\subsection{Systematic effects in our data?}
\begin{figure}
\begin{center}
\epsfig{figure=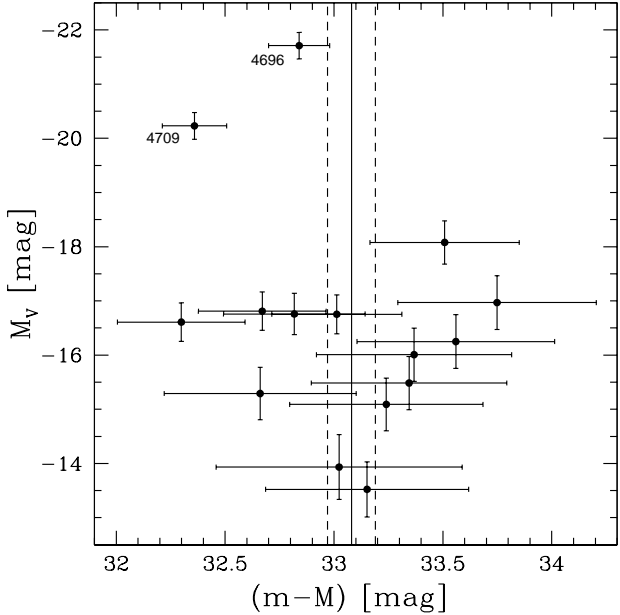,width=8.6cm}\\
\end{center}
\caption[]{\label{dMV}Distance modulus $(m-M)$ of the investigated galaxies plotted vs. their absolute magnitude
$M_V$. The solid vertical line indicates the mean distance of all galaxies. The dashed lines indicate the error
range of the mean distance. 4 of the 15 investigated galaxies fall outside of this error range.}
\end{figure}

Are there significant systematic selection effects present in our data?\\
In our final data set, we include three galaxies with S/N smaller than 3, whose mean 
distance is 46.4 $\pm$ 5 Mpc. Excluding them from the sample lowers the average 
distance by 1.7 Mpc, which is below significance. This does not indicate the presence of
a strong selection effect as mentioned in the former section. An additional test
is whether we can see a correlation between $(m-M)$ and $M_V$ for the investigated galaxies. Assuming
that we probe galaxies at the same distance, then the above mentioned selection effect would
move the faintest galaxies observed to closer distances. Fig.~\ref{dMV} shows such a plot. There
is no dependence between $(m-M)$ and $M_V$ within our error ranges. If a systematic
effect is present, it is negligible compared to the measurement uncertainty.\\
The same plot serves to check for an overall distance difference between giant and dwarf galaxies.
Separating the investigated galaxies by magnitude into giants ($M_V \le -17$ mag) 
and dwarfs ($M_V \ge -17$ mag) (Hilker et al. \cite{Hilker03}), we get a mean distance of 39 $\pm$ 6 Mpc
for the giants and 41.9 $\pm$ 2.3 Mpc for the dwarfs, i.e. there is no significant distance difference
between giants and dwarfs.\\
\subsection{Distance to NGC 4709, Cen30$-$Cen45 separation}
\label{separation}
\begin{figure}
\begin{center}
\epsfig{figure=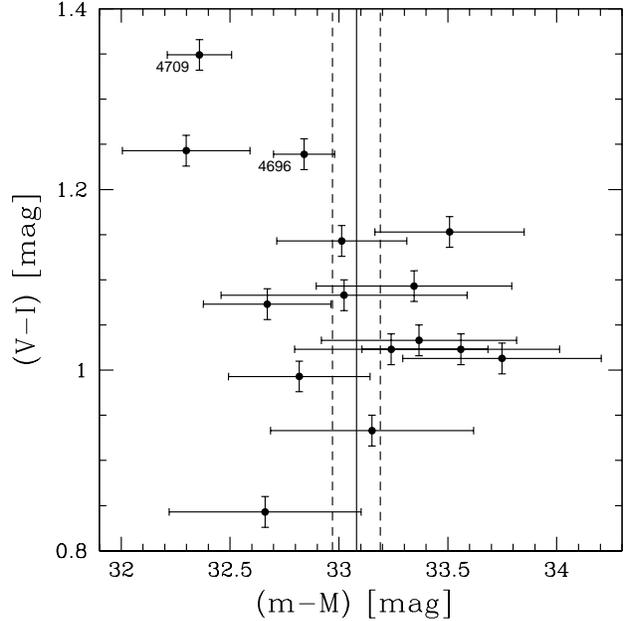,width=8.6cm}\\
\end{center}
\caption[]{\label{dmVI}Distance modulus $(m-M)$ of the investigated galaxies plotted vs. their colour $(V-I)$. 
Vertical lines as in Fig.~\ref{dMV}.}
\end{figure}

In Fig.~\ref{dmVI}, we plot $(m-M)$ vs. $(V-I)$. Here, a correlation would hint at systematic
errors in derivation of $\overline{M}_{\rm I}$ from $(V-I)$, under the assumption that all investigated
galaxies are at the same distance. Indeed, there seems to be a trend towards smaller
distances for redder galaxies. This trend is, however, almost entirely defined by the reddest investigated galaxy
NGC 4709. Including the NGC 4709 data point, there is a linear relation between $(m-M)$ and $(V-I)$ with 
slope different
from zero with 3.4 $\sigma$ significance. Excluding it, the significance drops to 1.45 $\sigma$. 
For comparison,
excluding the bluest data point only slightly increases the significance to 3.7 $\sigma$.\\
The issue condenses to the
question: Is the small distance for NGC 4709, placing it 7.5 $\pm$ 3 Mpc closer to us than NGC 4696, 
due to an unusual stellar population causing a bias in deriving 
$\overline{M}_{\rm I}$ from $(V-I)$, or, is the small distance a real physical fact? 
A real small distance is 
supported by the investigation
of NGC 4709's and NGC 4696 's GCS. It reproduces both the absolute distance of both galaxies and 
their distance difference
derived by SBF, see Table~\ref{resultsgcs}. 
Theoretical stellar population models (see for example Fig.~\ref{visbftheo} or {\it MieskeI}), 
although they frequently
predict {\it offset} $\overline{M}_{\rm I}$ values compared to equation~(\ref{sbfrel}), do not allow for a strong 
{\it scatter} in $\overline{M}_{\rm I}$ at a given $(V-I)$ for red colours, which as well favours a real small
distance instead of a stellar population effect mimicking it.\\
We therefore believe that the scenario of NGC 4709 being at a closer distance than NGC 4696 
is more probable than NGC 4709's bright SBF magnitude being caused by unusual stellar 
populations. Note that the distance difference between NGC 4709 and NGC 4696 in combination with
their velocity difference is consistent with
the Hubble flow distortion caused by the Great Attractor, if one assumes that 
NGC 4696 with Cen30 is located at the GA's center. In the lower panels of Fig.~18 and 19 of {\it SBF II} 
the distortion of the Hubble
flow along the line of sight towards the GA is shown as a function of distance. There are two  
distance ranges with respect to the GA's center (at 43 Mpc) where a radial velocity of 4500 km/s is reached: 
one between approximately $-$10 and $-2$ Mpc in front of the GA and one at about $+$ 20 Mpc behind. 
The former range corresponds to 
the approximately sinusoidally shaped Hubble flow distortion pattern in front of the GA. 
The latter value corresponds to the undisturbed Hubble flow.\\
With the radial velocity difference of about 1500 km/s and the distance difference of 
about 7.5 $\pm$ 3 Mpc between
NGC 4696 and NGC 4709 one can calculate the time difference left until NGC 4709 reaches the same 
distance as NGC 4696, assuming
that the infall velocity does not change considerably over time. The result is 5 $\pm$ 2 Gyrs. This is a very large
value, which is not consistent with an ongoing merger scenario between Cen30 and Cen45 in which NGC 4709 
participates, as for example proposed by Churazov et al. (\cite{Churaz99}) based on an X-ray temperature map
of the central Centaurus cluster. However, Furusho et al. (\cite{Furush01}) have suggested based on improved
X-ray data that a major merger is not taking place right now but already occured several Gyrs ago. 
Our finding that NGC 4709 is falling into Cen30 but 
has not yet reached it yet then 
suggests that NGC 4709 has not been involved with a merger
event during the last few Gyrs, but might be the subject of the next merger to come in several Gyrs more.
Nevertheless, we cannot exclude that NGC 4709 has passed by NGC 4696 in the more distant past.\\
\subsubsection{Filamentary structure}
\label{filament}
It has been suggested by Churazov et al. (\cite{Churaz99}) that in the direction of the Centaurus cluster we are 
looking into a large scale filamentary structure. They bring this scenario forward in order to explain an unusually 
extreme
``$\beta$-problem'', i.e. a substantial disagreement between the energy-ratio per unit mass for galaxies to that in the
gas derived from X-ray temperatures and from galaxy velocity dispersion. Colberg et al. (\cite{Colber99}) have shown
in numerical simulations that clusters accrete matter from a few preferred directions, defined by filamentary 
structures, and that the accretion persists over cosmologically long times.\\
This scenario of a filamentary 
structure is supported by the radial distance difference of 7.5 $\pm$ 3 Mpc between 
NGC 4696 and NGC 4709 derived by us, compared to their small angular separation of about 0.25 degrees or 0.2 Mpc
projected distance.\\
Precise SBF-distances to more Centaurus cluster giants must be measured to prove a
filamentary structure along the line of sight
towards Centaurus. Related to this subject, in the next section an upper limit on the depth of 
the investigated Centaurus cluster portion
is derived, discussing as well the distance scatter observed by Tonry et al. ({\it SBF IV}).\\
\subsection{Depth of the Centaurus Cluster}
\label{discussiondepth}
As can be seen in Fig.~\ref{dmhist}, the measured distance scatter of our data is almost
equal to the mean single measurement uncertainty, allowing no derivation of a lower limit for
the Centaurus cluster's depth. We can derive an upper limit for the depth, though.
We apply the inequality $\frac{(n-1)^2 (\Delta x)^2}{\chi ^2_{1-\frac{\alpha}{2}}}\le {\sigma}^2 \le \frac{(n-1)^2 (\Delta x)^2}{\chi ^2_{\frac{\alpha}{2}}}$
to obtain the confidence interval for the real variance
$\sigma ^2$ of a distribution
with a measured variance $\Delta x ^2$. In this inequality, the error probability of the confidence
interval is 
given by $1- \alpha$. $n-1$ denotes the number of degrees of freedom, in our case $n-1=14$.
From tabulated $\chi^2$ values we find that the Centaurus cluster would have to be
radially extended over more than 10 Mpc to both sides in order to exclude with more than 95\% confidence 
a $\delta$-distribution for the distance of our sample galaxies.\\
We therefore derive a formal 
upper limit of $\pm$ 10 Mpc radial extension for the Centaurus cluster. The cluster's angular extension
on the sky of 6 degrees (Lucey et al. \cite{Lucey86}) corresponds to a diameter of 4.3 Mpc at 41.3 Mpc distance. 
This would make us expect a distance scatter of about 2 Mpc around the mean in case of a spherical cluster shape.
I.e. we are only sensitive to a cigar-shape with
the major axis at least five times larger than the minor axis, leaving enough space for a possible
filamentary structure as proposed in Sect.~\ref{filament}.\\
Tonry et al. ({\it SBF IV}) measure SBF-distances for a total of 8 Centaurus cluster giants. 
Their mean distance is 33.7 Mpc with a rms-scatter of 3.9 Mpc and a mean measurement error
of 5.3 $\pm$ 1.2 Mpc. The distance scatter is even smaller than the measurement error at about 1 $\sigma$ 
significance. I.e. with a measurement accuracy about 30 \% better than ours, Tonry et al. do not find a 
significant radial extension of the Centaurus cluster. However, Tonry et al. only have a sample of 8 galaxies 
instead
of the 15 galaxies in our sample, which weakens the statistical significance by about the same factor of 30\%.\\
Therefore, the upper limit for the cluster's depth derived by us cannot be improved with
their data. Besides, the selection effect inherent in Tonry's data will generally decrease the distance scatter,
as those objects for which the observational errors give a larger distance are more likely to be excluded from 
their survey.\\
\subsection{Overestimation of the Great Attractor mass?}
\label{cenga}
The very large GA mass of almost 10$^{16}$ M$_{\rm sun}$ derived in {\it SBF II} was partially a consequence
of the large peculiar velocities observed by Tonry et al. for the Centaurus cluster galaxies, which
implied a very strong gravitational pull into the GA. Using
their mean Cen30 distance of 32.0 $\pm$ 1.8 Mpc and $H_0=71$ km/s/Mpc as derived by the WMAP team,
the Hubble flow velocity for Cen30 would be 2270 $\pm$ 180 km/s, 900 km/s smaller than the 
mean heliocentric radial velocity 3170 $\pm$ 170 km/s 
of the Stein et al. (\cite{Stein97}) sample of early type Centaurus cluster galaxies. Taking into account
the 300 km/s peculiar velocity of the Local Group towards Centaurus ({\it SBF II}), the peculiar velocity
of Cen30 even becomes 1200 $\pm$ 270 km/s. However, already Blakeslee et al. (\cite{Blakes02}) noted
that the distances at the faint limit of Tonry et al.'s survey are systematically underestimated by about
0.3 mag suggesting smaller peculiar velocities for Centaurus, a bias which had not been taken 
into account for the derivation of the GA mass in {\it SBF II}.\\
With our new distance value of 41.8 $\pm$ 3.4 Mpc for Cen30, the Hubble flow velocity of Cen30 
becomes 2970 $\pm$ 280 km/s, yielding a not significant Cen30 peculiar velocity of 200 $\pm$ 330 km/s.
Even when including the Local Group peculiar motion, the peculiar velocity of Cen30 is only 500 $\pm$ 340
km/s.\\
Our results thus imply that the Hubble flow distortion in the direction of the Centaurus cluster is smaller
by a factor of 2 or more compared to the distortion obtained by Tonry et al. and might even be negligible.
There are two interpretations for this:
First, the mass of the GA has been significantly 
overestimated in {\it SBF II}. Second, the Centaurus cluster falls into the Great Attractor with a
vector almost perpendicular to the line of sight.
 Then, the
radial component of its peculiar velocity would be comparably small. 
The distance of 41.8 $\pm$ 3.4 Mpc to 
Cen30 derived by us is very close to the GA distance 
of 43 $\pm$ 3 Mpc derived in ({\it SBF II}).
In addition, Tonry and collaborators find that the Centaurus
galaxies pass above the GA by about 15 degrees, which corresponds to about 10 Mpc at the Centaurus 
cluster distance. Both findings
are consistent with the second interpretation.
Note, however, that the distance underestimation for galaxies at the distance limit of Tonry's survey
 would probably increase the GA distance if
 corrected for.\\
To determine whether and to what degree either a mass overestimation of the GA or a mainly
tangential infall of the Centaurus cluster into the GA are responsible for the much smaller Hubble flow distortion 
of Centaurus cluster galaxies
observed by us compared to {\it SBF II}, high resolution SBF-distances to
galaxies in a much larger sky region than covered by us would be necessary.\\
\subsection{Backside infall?}
It has been claimed (Dressler \& Faber \cite{Dressl90}) and disclaimed (Mathewson et al. \cite{Mathew92}) 
that in the direction
of the GA there is an anticorrelation between redshift and distance. Such a back- and frontside 
infall pattern would be
 expected if the galaxies in front of the GA are drawn away from us, while the galaxies on the back side 
of the GA are drawn towards us (both falling into the GA). Can we verify such a behaviour from our data? 
Looking at 
Fig.~\ref{raddm} shows that there is certainly no correlation between redshift and distance, but
 neither there is any evidence for a significant anticorrelation. 
The mean radial velocity of the 4 galaxies
whose distance is larger than the mean distance is 3280 km/s. For the 7 galaxies
which are closer it is 3330 km/s. Thus, there is no correlation between distance and velocity when separating the
sample according to distance.
The distance modulus corresponding to
the mean distance of our Cen30 sample is 33.11 $\pm$ 0.17 mag, for our Cen45 sample it is 32.84 $\pm$ 
0.29 mag. When separating the sample according to radial velocity, there is a weak anticorrelation 
between redshift and distance, but not at a significant level. Only if one considers NGC 4696 and NGC 4709 alone,
an anticorrelation at 2.3 $\sigma$ significance is seen. This has already been discussed 
in Sect.~\ref{separation} and been found consistent with the Hubble flow distortion. However, only
two data points forming an anticorrelation, and this not even at 3 $\sigma$, 
is not sufficient to claim such a phenomenon for the entire cluster population.\\
Based on our data we can 
therefore not find any evidence for back- or frontside infall. Note however that due to the large 
distance uncertainties 
and the low number of galaxies we 
could only detect very pronounced infall patterns which extend over several tens of Mpc along the line of sight.
Besides, the sky position of the Great Attractor is deplaced by about 15 degrees with respect to the 
Centaurus cluster according to {\it SBF II}. The Centaurus galaxies might therefore not experience the GA's 
maximum gravitational pull.\\
\section{Summary and conclusions}
We have presented $I$-band SBF-measurements for 15 early type Centaurus cluster galaxies in
the magnitude range $19.6>V>11.5$ mag, 3 giant and 12 dwarf galaxies. 
The measurements were made on deep photometric data obtained
with VLT FORS1 in the $I$-band in 7 fields with a seeing between 0.4 and 0.6 $''$. 
The following results
were obtained:\vspace{0.2cm}\\
\indent 1. The mean distance of our investigated galaxies is 41.3 $\pm$ 2.1 Mpc. This corresponds
to a distance modulus of 33.08 $\pm$ 0.11 mag and places the Centaurus cluster at the same distance 
as the Great Attractor
(see Tonry et al. \cite{Tonry00}). We find that our Centaurus cluster distance is about 0.5 mag higher
than Tonry's value (Tonry et al. \cite{Tonry01}). This is explained by the fact that the sensitivity
limit of Tonry's SBF-survey is reached at about the Centaurus distance and that therefore the 
galaxies whose observational errors place them further away are less likely to enter their survey.\vspace{0.1cm}\\
\indent 2. Splitting our data according to their measured redshifts into Cen30 and Cen45, we obtain 
a distance difference $(m-M)_{\rm Cen45}$ $-$ $(m-M)_{\rm Cen30}$=
$-$0.27 $\pm$ 0.34 mag. 
This rules out both components being separated by their
Hubble flow distance but still allows for a wide range of separation, including no separation
at all. We do find a
significant separation of $(m-M)_{\rm NGC 4709}$ $-$ $(m-M)_{\rm NGC 4696}$=
$-$0.48 $\pm$ 0.21 mag between the two dominant giants of Cen45 and Cen30, supported by the measured
turn-over magnitudes of their respective Globular Cluster systems and supporting a scenario
with Cen45 being a subgroup falling into but not having reached Cen30 yet. This
scenario is found to be consistent with the proposed large scale filament along the line of sight
towards Centaurus (Churazov et al. \cite{Churaz99}) and the Hubble flow distortion caused by the Great Attractor 
(Tonry et al. \cite{Tonry00}).\vspace{0.1cm}\\
\indent 3. The Hubble constant $H_{\rm 0}$ is determined to be 83.0 $\pm$ 8.3 km/s/Mpc for our Cen30 
sample taking into account the 
peculiar motion of the Local Group into the direction of the Centaurus cluster.\vspace{0.1cm}\\
\indent 4. The peculiar velocity of Cen30 with respect to an undisturbed Hubble flow is 500 $\pm$ 340 km/s
when taking into account the 
peculiar motion of the Local Group into the direction of Centaurus, and only 200 $\pm$ 330 km/s
when not. This means a much smaller Hubble flow distortion in the direction of Centaurus than
previously obtained by Tonry et al. (\cite{Tonry00}) and implies that the GA mass estimate
by Tonry et al. may be too high and/or that the Centaurus cluster falls into the GA almost
perpendicularly to the line of sight.\vspace{0.1cm}\\
\indent 5. We cannot place lower limits on the Centaurus cluster depth from our data, as the measured
distance scatter of our sample (8.1 Mpc) is almost equal to the mean single measurement uncertainty (7.3 Mpc). 
We can place an upper limit of $\pm$ 10 Mpc radial depth, corresponding to a five 
times larger radial than tangential extension.\vspace{0.1cm}\\
\indent 6. We find no significant anticorrelation between redshift and 
distance for our data. Such a pattern would be expected in case of a backside infall into
the Great Attractor. Our number counts are too low to make more definite statements, 
especially given the large distance uncertainty of our data.\vspace{0.2cm}\\
\noindent We conclude that the deep and high resolution SBF-measurements presented here prove that
the SBF-method allows a precise measurement of cluster distances with ground 
based imaging out to 40 Mpc and
beyond. The 2 highest S/N measurements for NGC 4696 and NGC 4709 show that distance accuracies better
than 10\% can be achieved easily if the sampling area for SBF measurement is large enough.
We have shown that it is essential to obtain sufficiently high S/N SBF data such that
biases like the selection effect in distance determinations at the sensitivity limit of
Tonry's SBF survey (Tonry et al \cite{Tonry97}) do not occur. To better investigate the structure
of the Centaurus cluster with respect to its possible filamentary form and its relation to the
Great Attractor, a deep and wide field survey of the 
entire cluster region would be needed.\\
\label{conclusions}
\acknowledgements
The authors wish to thank the referee Dr. H. Jerjen for his very valuable comments which helped a 
great deal to improve the paper. SM was supported by DAAD PhD grant Kennziffer D/01/35298. 
The authors would like to thank the ESO user support group and the ESO science operation for having 
carried out the 
programme in service mode.\\

\enddocument